\documentclass[11pt, a4paper]{article}%
\usepackage{amsmath}%
\usepackage{amsfonts}%
\usepackage{amssymb}%
\usepackage{color}
\usepackage{xypic}
\usepackage{hyperref}
\hypersetup{colorlinks=red,citecolor=red,urlcolor=red,linkcolor=red}

\newtheorem{theorem}{Theorem}

\newtheorem{corollary}[theorem]{Corollary}

\newtheorem{definition}[theorem]{Definition}
\newtheorem{example}[theorem]{Example}

\newtheorem{lemma}[theorem]{Lemma}
\newtheorem{notation}[theorem]{Notation}

\newtheorem{proposition}[theorem]{Proposition}
\newtheorem{remark}[theorem]{Remark}

\newenvironment{proof}[1][Proof]{\noindent\textbf{#1.} }{\ \rule{0.5em}{0.5em}}

\newcommand{\NN}{\mathbb N}

\def\N{\mathbb{N}}

\def\Q{\mathbb{Q}}

\def\K{\mathbb{K}}
\def\L{\mathbb{L}}

\def\cF{\mathcal{F}}
\def\cB{\mathcal{B}}
\def\fJ{\mathfrak{J}}

\def\square{\ \rule{0.5em}{0.5em}}

\begin{document}

\title{A linear algebra approach to the differentiation index
of generic DAE systems }

\author{Lisi D'Alfonso\thanks{Partially supported by the following
research grants: UBACyT X112 (2004-2007) and CONICET PIP 02461/01.} \\[0.1cm]
{\textit{\normalsize Departamento de Matem\'atica, Facultad de Ciencias
Exactas y Naturales,}}\\
{\normalsize{and} \textit{\normalsize Departamento de Ciencias Exactas, Ciclo B\'asico Com\'un,}}\\
{\textit{\normalsize Universidad de Buenos Aires, Ciudad Universitaria, 1428
Buenos Aires, Argentina}}
\\[0.2cm]\and
Gabriela Jeronimo$^*$\thanks{Partially supported by CONICET PIP 5852/05, UBACyT
X847 (2006-2009) and ANPCyT PICT 2005 17-33018.}, Pablo Solern\'o$^*$
\\[0.1cm]
{\textit{\normalsize Departamento de Matem\'atica, Facultad de Ciencias
Exactas y Naturales}}\\
{\textit{\normalsize Universidad de Buenos Aires, Ciudad Universitaria, 1428
Buenos Aires, Argentina}}\\{\normalsize and  CONICET,  \textit{Argentina}}
\\[0.2cm]
{\normalsize E-mail: lisi@dm.uba.ar, jeronimo@dm.uba.ar,
psolerno@dm.uba.ar}}

\maketitle

\begin{abstract}
The notion of differentiation index for DAE systems of arbitrary
order with generic second members is discussed by means of the study
of the behavior of the ranks of certain Jacobian associated
sub-matrices. As a by-product, we obtain upper bounds for the
regularity of the Hilbert-Kolchin function and the order of the
ideal associated to the DAE systems under consideration, not
depending on characteristic sets. Some quantitative and algorithmic
results concerning differential transcendence bases and induced
equivalent explicit ODE systems are also established.
\end{abstract}

\section{Introduction}

This paper is devoted to the study, mainly from a quantitative point of view,
of  differential algebraic equation (DAE) systems of the form:
\[
(\Sigma)\ :=\ \left\{
\begin{array}
[c]{ccl}%
f_1(X, U) &=& \dot{X}_1 \\
& \vdots & \\
f_n(X, U) &=& \dot{X}_n \\
g_1(X, U, \dot{U},\ldots ,U^{(e_1)}) &=& Y_1 \\
&\vdots &\\
g_r(X, U, \dot{U},\ldots,U^{(e_r)}) &=& Y_r \\
\end{array}
\right. ,
\]
where $f_1,\dots, f_n$ are polynomials in the $n+m$ variables
$X:=X_1,\ldots ,X_n$, $U:=U_1,\ldots ,U_m$, and, for every $1\le j
\le r$, $g_j$ is a polynomial in the $n+(e_j+1)m$ variables $X$, $U$
and the derivatives ${U}^{(i)}:={U}^{(i)}_1,\ldots ,{U}^{(i)}_m$,
$1\le i\le e_j$, with coefficients in a differential field $k$ (for
instance $k:=\mathbb{Q}$, $\mathbb{R}$, $\mathbb{C}$,
$\mathbb{Q}(t)$, etc.). The constants $e_j\in \mathbb{N}_0$ denote
the order of the respective equation $g_j$ in the variables $U$. The
variables $Y:=Y_1,\ldots ,Y_r$ are a new set of differential
indeterminates which can be viewed as parameters (while the
variables $X$ and $U$ are the unknowns of the system). So, it is
quite natural to extend the ground field $k$ to the differential
field $\mathbb{L}:=k\langle Y\rangle $ (i.e. the smallest field
containing $k$ and all the successive derivatives of $Y$) and
consider our input system also as a system over $\mathbb L$. Even if
we do not assume, as customarily, a differential $0$-dimensional
situation ($r$ may be strictly smaller than $m$), we will suppose
that the last $r$ equations are ``independent" in a suitable natural
way defined in Section \ref{basic}.

DAE systems like $(\Sigma)$ can be regarded from several points of
view: for instance, this kind of systems arises in Control Theory
(see for instance \cite{Fliess} and \cite[Section
4]{campbell-gear}); they may also be interpreted as the equations
defining the graph of a differential morphism (see \cite{MS}). The
system $(\Sigma)$ may be viewed as a family of usual polynomial DAE
systems where the second member parametrizes the family and takes
arbitrary values outside a suitable proper algebraic Zariski closed
set (see \cite[Section 5.2]{sadikb}). In this last sense we say that
the system $(\Sigma)$ is \emph{generic}.

We will focus on several basic topics concerning the system $(\Sigma)$,
including the number of differentiations  that suffice to obtain explicit
equations, a description of all the relations of a prescribed order that all
the solutions must verify and the number of initial conditions that can be
arbitrarily fixed. All these aspects have been studied extensively during the
last two centuries  beginning with two posthumous articles by Jacobi
(\cite{jacobi} and \cite{jacobi2}). The present paper (as well as most of the
previous ones on the subject) may be considered as a modern approach to the
work done by Jacobi in these remarkable and not sufficiently known papers.

The main notion we will consider is the \emph{differentiation index}, which is
a well known and important invariant associated to a DAE system (customarily
for first-order and $0$-dimensional systems).

There are many different, not always equivalent, definitions of differentiation
indices (see for instance \cite{brenan}, \cite{rabier}, \cite{campbell-gear},
\cite{Fliess}, \cite{reid}, \cite{sit}, \cite{kunkel}, \cite{levey},
\cite{seiler}, \cite{pantelides} and \cite{poulsen}). Here we are mainly
interested in the so-called \emph{global differentiation index} (see
\cite[Section 2.2]{brenan}). Roughly speaking, the differentiation index
represents the minimum number of times that all or part of a given DAE system
must be differentiated in order to obtain an \textit{equivalent explicit
ordinary differential equation (ODE) system}; in other words, the number of
differentiations required to determine the derivatives of a certain order of
the unknowns as continuous functions of derivatives of lower order of the
unknowns (see \cite[Definition 2.2.2]{brenan}). In some sense, the index can be
regarded as a measure of the complexity of the DAE system. From the theoretical
point of view it represents the distance between the given  system and another
one for which an existence and uniqueness theorem holds (see \cite{rabier}). On
the other hand, from the point of view of its numerical resolution, it is
closely related to the condition number of the iteration matrix in the implicit
Runge-Kutta method (see \cite[Theorem 5.4.1]{brenan}).

In this paper we give a precise algebraic definition of a differentiation index
for DAE systems as ($\Sigma$), \emph{not necessarily of first order nor
$0$-dimensional} (see Definition \ref{indice} below), by means of certain
stationary properties of the ranks of suitable Jacobian sub-matrices which are
proved in Section \ref{linearjacobian}. Another equivalent definition of this
index, in terms of a quite natural filtration given by the successive
differentiation of the input equations, is implicitly contained in Theorem
\ref{sadik} below. In particular, this last formulation shows how the
differentiation index uncovers constraints that every solution must satisfy.
This approach is closely related to the classical algorithmic definitions of
the differentiation index by means of iterated \emph{prolongations}
(differentiations) and \emph{projections} (eliminations) (see \cite{rabier},
\cite{cartan}, \cite{kuranishi} and \cite{reid}).


Our theoretical  approach has the advantage of  leading to  a polynomial time
algorithm which computes the index of the input system $(\Sigma)$  by simple
comparison of ranks of Jacobian matrices (see Section \ref{issues}). For
previous work related to the computation of differentiation indices of DAE
systems, we refer the reader to  \cite{cartan}, \cite{kuranishi},
\cite{rabier}, \cite{reid}, \cite{sit}, \cite{pantelides}, \cite{poulsen},
\cite{thomas} and \cite{lamour}.

As we said before, the notion of differentiation index is closely
related to the possibility of writing certain derivatives of the
unknowns in terms of derivatives of lower order of these unknowns.
Unfortunately, in general, this cannot be done using \textit{only}
the algebraic relations induced by the original equations. For
example, if all the equations have order one, this nice situation,
which is usually ensured by the Theorem of Implicit Functions,
corresponds exactly to those systems whose associated index is $0$.
However, in the general case, by successive differentiations (as
many as the index) we can always obtain such a situation. Evidently,
the new explicit system comes out of the frame of the polynomial (or
even rational) systems. Under certain additional conditions on the
system $(\Sigma)$, we are able to compute an implicit ODE system
which is equivalent to $(\Sigma)$ (see Section \ref{ODE}); in
particular, in the case of a first-order system, we give an implicit
but simple polynomial way to describe it, distinguishing the
variables by their interrelations (namely, ``free variables",
``implicit variables", etc.). Moreover, we can estimate degree and
order upper bounds for the implicit equations involved in the
equivalent ODE systems (see Subsection \ref{grados}) and give an
algorithm to compute them with polynomial complexity in terms of an
intrinsic parameter related to the geometric degree of suitable
associated algebraic varieties (Proposition \ref{algoritmo}).


Our method also allows us to give a new upper bound for the
\emph{regularity} of the \emph{Hilbert-Kolchin function} (or
\emph{differential Hilbert function}) associated to the DAE system
$(\Sigma)$. The Hilbert-Kolchin function is introduced in
\cite[Chapter II]{kol} in order to estimate, for each non-negative
integer $i$, the degree of freeness of the first $i$-derivatives of
the unknowns modulo the relations induced by the input equation
system (see also Section \ref{Hilbertfunction} below for a precise
definition in our case).

As it happens for the classical Hilbert function associated to
homogeneous polynomial ideals (see for instance \cite[Chapter
11]{atiyah}), the Hilbert-Kolchin function becomes a well defined
polynomial for sufficiently big arguments $i$ (in the ordinary
differential setting, this polynomial is extremely simple since its
degree is at most $1$). The \emph{regularity} of the Hilbert-Kolchin
function is defined to be the first non-negative integer from where
the function and the polynomial coincide. It is well known that this
regularity can be exactly described in terms of the orders of the
elements in a characteristic set associated to any orderly ranking
(see the proof of \cite[Chapter II, Section 12, Theorem 6]{kol} or
\cite[Theorem 3.3]{carra}). In this article (see Theorem
\ref{hilbert constante} below) we show that $\max\{1,e_1,\ldots,
e_r\} -1$ is an upper bound for the regularity of the Hilbert
function for the system $(\Sigma)$ over the field $\mathbb L$
(notice that no characteristic sets are involved in the bound). In
particular, if $(\Sigma)$ is a first-order system, the regularity of
the Hilbert-Kolchin function is $0$; in other words, the
Hilbert-Kolchin function and the associated polynomial coincide for
all $i$.

As further  consequences of our techniques  we deduce B\'ezout-type bounds for
 the differentiation index and the order of  the input system $(\Sigma)$ in
terms of the orders of its defining equations (see Remark \ref{inequalities}
and Subsection \ref{orders}). A similar bound for the differentiation index may
also be obtained by rewriting methods as in \cite[Section 5.2]{sadikb} (see
also \cite[Section 3]{Fliess} for an analogous bound for $0$-dimensional
first-order systems). Concerning  the order, we recover Ritt's differential
analogue to B\'ezout's Theorem (\cite[Ch.~VII, p.135]{ritt}).  A more precise
bound  for the order of a $0$-dimensional  system was conjectured by Jacobi
(see \cite{jacobi}) and proved in \cite{kondra}  under  additional hypotheses
which are met in our situation (see also \cite{ritt2}, \cite{ollisad}). We
point out that  a convenient refinement of our approach may be applied  to
obtain Jacobi-type bounds for both the order and the differentiation index,
which will be the subject of a forthcoming paper.

%

\bigskip

The paper is organized as follows: basic definitions and notations
are introduced in Section \ref{basic0}. Section \ref{linearjacobian}
is devoted to the study of the behavior of an integer sequence which
is strongly related to the ranks of suitable Jacobian sub-matrices
of the input equations and their successive derivatives. In Section
\ref{seccind} we give a precise algebraic definition of the
differentiation index based on the behavior of this sequence, and an
equivalent description of it in terms of the variety of constraints
(Subsection \ref{sectionindex}). Our results on the Hilbert-Kolchin
function of the differential ideals associated with the considered
DAE systems are presented in Subsection \ref{Hilbertfunction}: we
estimate the regularity of the Hilbert-Kolchin function in
Subsection \ref{secregularity} and the order of the induced
differential ideal in terms of the orders of the equations in the
given system in Subsection \ref{orders}; further, a first result
related to equivalent explicit ODE systems is given for the
differential $0$-dimensional case in Subsection \ref{ODEdim0}. This
result is refined in Section \ref{ODE} by introducing a more
accurate definition of the index, which is done in Subsection
\ref{sigma mono}, and generalized for first-order systems with
positive differential dimension in Subsection \ref{implicit}.
Finally, we present quantitative and algorithmic considerations
concerning these results in Subsection \ref{quantitative}.

\bigskip

\noindent \textbf{Acknowledgements\ }\  The authors thank Evelyne
Hubert (INRIA, Sophia Antipolis), Gustavo Massaccesi (Universidad de
Buenos Aires), Fran\c cois Ollivier (\'Ecole Polytechnique,
Palaiseau) and Alexandre Sedoglavic (Universit\'e de Lille) for
their helpful remarks.

\section{Preliminaries}
\label{basic0}
\subsection{Basic definitions and notations} \label{basic}

Let $k$ be a characteristic zero field equipped with a derivation
$\delta$ (for instance $k=\mathbb{Q}$, $\mathbb{R}$ or
$\mathbb{C}$ with $\delta:=0$, or $k=\mathbb{Q}(t)$ with the usual
derivation $\delta(t)=1$, etc.).

For an arbitrary set of (differential) indeterminates $Z:=
Z_1,\ldots ,Z_\alpha$ over $k$ we denote the $l$-th successive
derivative of a variable $Z_j$ as $Z_j^{(l)}$ (as customarily, the
first derivatives are also denoted by $\dot{Z_j}$); we write
$Z^{(l)}:=\{Z^{(l)}_1,\ldots,Z^{(l)}_\alpha\}$ and
$Z^{[l]}:=\{Z^{(i)},\ 0\le i\le l\}$. The (non-noetherian)
polynomial ring $k[Z^{(l)},\ l\in \mathbb{N}_0]$, called the ring of
\emph{differential polynomials}, is denoted by $k\{Z_1,\ldots
,Z_\alpha\}$ and its fraction field by $k\langle Z_1,\ldots
,Z_\alpha \rangle$. Given a finite set of (differential) polynomials
$H_1,\dots, H_\beta \in k\{ Z_1,\dots, Z_\alpha\}$, we write
$[H_1,\dots, H_\beta]$ to denote the smallest ideal of
$k\{Z_1,\dots, Z_\alpha\}$ stable under differentiation, i.e. the
smallest ideal containing $H_1,\dots, H_\beta$ and all their
derivatives of arbitrary order. The ideal $[H_1,\dots, H_\beta]$ is
called the \emph{differential ideal} generated by $H_1,\dots,
H_\beta$.

We deal with a particular class of differential algebraic equation
(DAE) systems:
\begin{equation} \label{sistema ampliado}
\left\{
\begin{array}
[c]{ccl}%
f_1(X, U) &=& \dot{X}_1 \\
& \vdots & \\
f_n(X, U) &=& \dot{X}_n \\
g_1(X, U, \dot{U},\ldots ,U^{(e_1)}) &=& Y_1 \\
&\vdots &\\
g_r(X, U, \dot{U},\ldots,U^{(e_r)}) &=& Y_r \\
\end{array}
\right.
\end{equation}
where $f_1,\dots, f_n$ are polynomials in the $n+m$ variables
$X:=X_1,\ldots ,X_n$, $U:=U_1,\ldots ,U_m$, and, for every $1\le j
\le r$,  $g_j$ is a polynomial in the $n+(e_j+1)m$ variables $X$,
$U$ and the derivatives ${U}^{(i)}:={U}^{(i)}_1,\ldots
,{U}^{(i)}_m$, $1\le i\le e_j$, with coefficients in the field $k$.
The constants $e_j\in \mathbb{N}_0$ denote the order of the
respective equation $g_j$ in the variables $U$, i.e., the order of
the highest derivative of a variable in $U$ appearing with non-zero
coefficient in this polynomial. The variables $Y:=Y_1,\ldots ,Y_r$
form a new set of indeterminates which we regard  as parameters
(while the variables $X$ and $U$ are the unknowns of the system). We
allow $n$ to be equal to zero (i.e. no variables $X$ appear in the
system).

In addition, we will assume that the polynomials $g_i$ are
\emph{differentially algebraically independent over $k$} as elements
of the fraction field $\mbox{\rm Frac}(k\{Y, X, U\}/[f_1 -
\dot{X}_1, \dots ,f_n - \dot{X}_n])$, i.e. there is no non-trivial
differential relation involving the classes of the differential
polynomials $g_1,\ldots ,g_r$ over $k$. This assumption guarantees
that the differential ideal of $k\{Y, X, U\}$ associated to system
(\ref{sistema ampliado}) does not contain a non-zero differential
polynomial involving only variables $Y$.

For  any differential polynomial $g$ lying in a differential
polynomial ring $k\{ Z_1,\ldots ,Z_\alpha \}$ the following
recursive relations hold for the successive derivatives of $g$:
$$\begin{array}{rcl}
g^{(0)}&:=& g, \\
g^{(l)}&:=& \delta(g^{(l-1)})+\sum_{i,j} \dfrac{\partial
g^{(l-1)}}{\partial Z_j^{(i)}}Z_j^{(i+1)}, \quad \hbox{ for } l\ge
1, \qquad \end{array}$$ where $\delta(g^{(l-1)})$ denotes the
polynomial obtained from $g^{(l-1)}$ by applying the derivative
$\delta$ to all its coefficients (for instance, if $k$ is a field
of constants, this term is always zero).

Concerning the system (\ref{sistema ampliado}) we also introduce
the following definitions and notations:
\begin{itemize}
\item $\mathbb{L}$ denotes the fraction field $k\langle Y
\rangle$.
\item $e:=\max \{ 1, e_1,e_2,\ldots ,e_r\}$.
\item For every $l\in \mathbb{N}_0$ we set:
\begin{eqnarray*}
F_i^{(l)} &:=& f_i^{(l)}-X^{(l+1)}_i\, \in \, k[X^{[l+1]},U^{[l]}]
\qquad \quad i=1,\dots, n,\\
G_j^{(l)} &:=& g_j^{(l)}-Y_j^{(l)}\, \in \, k[Y^{[l]},X^{[l]},
U^{[l+e_j]}] \qquad j=1,\dots, r.
\end{eqnarray*}
\item For every $l\in \NN$, $A_l$ denotes the polynomial ring
$A_l:=\mathbb{L}[X^{[l]}, U^{[l]}]$ and $\Delta_l\subset
A_{l-1+e}$ the ideal generated by $F^{[l-1]}, G^{[l-1]}$ (observe
that the ideal $\Delta_l$ is contained in $A_{l-1+e}$ because the
orders of $F^{(l-1)}$ and $G^{(l-1)}$ are at most $l-1+e$). We
set $\Delta_0:=(0)$.
\item $\Delta:= [F, G]\subset \mathbb{L}\{ X, U\}$ is the
{differential} ideal generated by the polynomials $F:= F_1, \dots,
F_n$ and $G:= G_1, \dots, G_r$.

\item $\K$ is the differential field $k(X)\langle U \rangle$ with
the derivation induced by $\dot{X}_i = f_i(X, U)$, $i = 1,\dots,
n$.

\end{itemize}

The hypothesis on the differential algebraic independence of the
polynomials $g_j$ $(1\le  j \le r)$ easily implies that \textit{the ideal
$\Delta$ has differential dimension $m-r$ over $\mathbb{L}$} (see
for instance \cite[Proposition 12]{djs}). Furthermore, the map
\[\dot{X}_j  \mapsto  f_j, \ j=1, \dots, n, \quad Y_j \mapsto g_j, \
j=1,\dots, r,\quad \textrm{and}\quad U_j\mapsto U_j, \ j=1,\ldots,
m,
\]induces an isomorphism between the fraction field ${\rm
Frac}\left( \L\{ X, U\} /\Delta \right)$ and the differential field
$\K= k(X)\langle U\rangle$ (see \cite[Remark 7]{djs}).

In the sequel, in order to simplify notations, for any $g \in \L\{
X, U\}$, we will also write $g$ for its class in $\L\{ X, U\}
/\Delta$ or in its fraction field $\K$ (the ring where we consider
the object will be clear by the context). In the same way, $\dot g$
will denote either the derivative of $g$ in $\L\{ X, U\}$ or its
derivative as an element of $\K$, and so on.

\subsection{Associated Jacobian sub-matrices}

Here, we introduce the Jacobian matrices and sub-matrices we will
deal with throughout the paper. {}From the considered
differential system (\ref{sistema ampliado}), we define a family
of sub-matrices constructed from the (infinite) Jacobian matrix
associated to the (infinitely many) polynomials $F^{(l)}$ and
$G^{(l)}$ with respect to the (infinitely many) variables
$X^{(j)}$ and $U^{( j)}$.
\begin{definition} \label{defimatricespi}
For each $k\in\N$ and $i\in \mathbb{N}_{\ge e-1}$ (i.e. $i\in
\mathbb{Z}$ and $i\ge e-1$), we define the matrix \
$\mathfrak{J}_{k,i}\in \K^{k(n+r)\times k(n+m)}$ as follows:
\[ \mathfrak{J}_{k,i}:= \left(
\begin{array}{ccccccc}
\frac{\partial F^{(i-e+1)}}{\partial {X}^{(i+1)}} & \frac{\partial
F^{(i-e+1)}}{\partial {U}^{(i+1)}} & 0 & 0 & \cdots  & 0 & 0
\\[1.5mm]
\frac{\partial {G}^{(i-e+1)}}{\partial {X}^{(i+1)}} & \frac{\partial
{G}^{(i-e+1)}}{\partial U^{(i+1)}} & 0 & 0&
\cdots  & 0 & 0 \\[1.5mm]
\vdots  & \vdots  & \vdots  & \vdots  & \ddots  & \vdots & \vdots \\[1.5mm]
\frac{\partial F^{(i-e+k)}}{\partial {X}^{(i+1)}} & \frac{\partial
F^{(i-e+k)}}{\partial U^{(i+1)}} & \frac{
\partial F^{(i-e+k)}}{\partial X^{(i+2)}} & \frac{\partial F^{(i-e+k)}}{\partial U^{(i+2)}} &
\cdots
 & \frac{\partial F^{(i-e+k)}}{\partial X^{(i+k)}}& \frac{\partial F^{(i-e+k)}}{\partial
 U^{(i+k)}}\\[1.5mm]
\frac{\partial G^{(i-e+k)}}{\partial {X}^{(i+1)}} & \frac{\partial G^{(i-e+k)}}{\partial U^{(i+1)}} & \frac{%
\partial G^{(i-e+k)}}{\partial X^{(i+2)}} & \frac{\partial G^{(i-e+k)}}{\partial U^{(i+2)}} & \cdots  &
\frac{\partial G^{(i-e+k)}}{\partial X^{(i+k)}}&
\frac{\partial G^{(i-e+k)}}{\partial U^{(i+k)}}%
\end{array}
\right).
\]

\end{definition}

In other words, $\mathfrak{J}_{k,i}$ is the Jacobian matrix of
the polynomials \[F^{(i-e+1)}, G^{(i-e+1)}, \dots, F^{(i-e+k)},
G^{(i-e+k)}\in \L[X^{[i-e+k+1]}, U^{[i+k]}]\] with respect to the
variables $X^{(i+1)}, U^{(i+1)}, \dots, X^{(i+k)}, U^{(i+k)}$,
where the entries are regarded as elements in $\K$.

Observe that the block triangular form of $\mathfrak{J}_{k,i}$
follows from the fact that the differential polynomials
$F^{(i-e+l)}$ and $G^{(i-e+l)}$ have order bounded by $i+l$.
Hence, their derivatives with respect to the variables
$X^{(i+j)}$ and $U^{(i+j)}$ are identically zero for $j\ge l+1$.

\bigskip

The matrices $\mathfrak{J}_{k,i}$ are strongly related with some
algebraic facts concerning the (algebraic) ideals $\Delta_l$
introduced in the previous subsection:

\begin{proposition} \label{equival}
Let $k\in \N$ and $i\in \mathbb{N}_{\ge e-1}$. Then:
\begin{enumerate}
\item[(i)] The transcendence degree of the field extension
\[{\rm Frac}(A_i/(\Delta_{i-e+1+k}\cap A_i)) \hookrightarrow {\rm
Frac}(A_{i+k}/\Delta_{i-e+1+k})\] equals the dimension of the kernel
of $\mathfrak{J}_{k,i}$.
\item[(ii)] The following identity holds:
$$
{\rm trdeg}_\L ({\rm Frac}(A_i/\Delta_{i-e+1+k}\cap A_i))  =
(m-r)(i+1) + e(n+r) - \dim_\K({\rm ker}(\fJ_{k,i}^t)),$$ where
$\fJ_{k,i}^t$ denotes the usual transpose of the matrix $\fJ_{k,i}$.
\end{enumerate}
\end{proposition}

\noindent{\textbf{Proof.}} In order to prove (i) we follow closely
the proof of \cite[Proposition 16]{djs} (see also \cite[Theorem
10]{MS}): since the polynomials $F_i^{(p)}$ and $G_j^{(p)}$ have
order $p+1$ and $p+e_j$ respectively, we conclude that for $i>e-1$,
the polynomials $F^{[i-e]}$ and $G^{[i-e]}$ belong to the ring
$A_i$. In any case (even if $i=e-1$), we may consider ${\rm Frac}
(A_{i+k}/\Delta_{i-e+1+k})$ as the fraction field of the domain
$$R:=K[X^{(i+1)},U^{(i+1)},\ldots
,X^{(i+k)},U^{(i+k)}]/(F^{(i-e+1)},G^{(i-e+1)},\ldots
,F^{(i-e+k)},G^{(i-e+k)}),$$ where $K$ denotes the field ${\rm
Frac}(A_i/(\Delta_{i-e+1+k}\cap A_i))$.

Then, the transcendence degree we want to compute is the dimension
over the field ${\rm Frac}(R)$ (or equivalently, over ${\rm
Frac}(A_{i+k}/\Delta_{i-e+1+k})$) of the kernel of the Jacobian
matrix associated to the $K-$algebra $R$ (see for instance
\cite[Chapter VI, \S1, Theorem 1.15]{kunz}). This Jacobian matrix is
exactly the matrix $\mathfrak{J}_{k,i}$. To finish the proof of (i),
it suffices to show that the dimension of both kernels (namely, over
${\rm Frac}(R)$ and over $\mathbb{K}$) of $\mathfrak{J}_{k,i}$ is
the same.

Let us observe that the entries of $\mathfrak{J}_{k,i}$ can be
regarded as polynomials in the ring $k[X,U^{[i+k]}]\subset
\mathbb{K}$, which is isomorphic to
$k[Y^{[i-e+k]},X^{[i+k]},U^{[i+k]}]/(F^{[i-e+k]},G^{[i-e+k]})$.
After tensoring by $\mathbb{L}$, this last ring becomes
$A_{i+k}/\Delta_{i-e+1+k}$. Now, since $\mathbb{L}\cap \Delta=0$,
the rank of $\mathfrak{J}_{k,i}$ is preserved after that tensoring
and so, its rank over the fraction field of $k[X, U^{[i+k]}]$ (and
therefore, over $\mathbb{K}$) is equal to its rank over the fraction
field of $A_{i+k}/\Delta_{i-e+1+k}$ (namely, ${\rm Frac}(R)$). This
finishes the proof of (i).

\medskip

For part (ii), it is easy to see that the polynomials
$F^{(p)},G^{(p)}$ form a regular sequence (see for instance
\cite[Corollary 9]{djs}) and hence, the field extension
$\mathbb{L}\hookrightarrow \textrm{Frac}(A_{i+k}/\Delta_{i-e+1+k})$
has transcendence degree equal to the number of variables in
$A_{i+k}$ minus the number of generators of the ideal
$\Delta_{i-e+1+k}$, that is
$(i+k+1)(n+m)-(i-e+k+1)(n+r)=(m-r)(i+k+1)+e(n+r)$. The result
follows by considering the tower of fields
\[\mathbb{L}\hookrightarrow
\textrm{Frac}(A_i/\Delta_{i-e+1+k}\cap A_i)\hookrightarrow
\textrm{Frac}(A_{i+k}/\Delta_{i-e+1+k})\] and part (i), noticing
that $\dim_{\mathbb{K}} {\rm ker} (\fJ_{k, i}) =
k(m-r)+\dim_{\mathbb{K}} {\rm ker}(\fJ_{k, i}^t)$. \square

\begin{remark} \label{k0}
The equality in part (ii) of Proposition \ref{equival} also holds
for $k=0$ defining $\dim_\K({\rm ker}(\fJ_{0,i}^t)):=0$ for $i\ge
e-1$. This follows from the fact that $\Delta_{i-e+1}\subset A_i$
is generated by a regular sequence, as in the proof of that
Proposition.
\end{remark}

\begin{remark} \label{noether}
Fix an index $i\ge e-1$. Note that the prime ideals
$(\Delta_{i-e+1+k}\cap A_i)_{k\in \mathbb{N}_0}$ form an
increasing chain of ideals. Since $A_i$ is a noetherian ring, this
chain must be stationary for $k$ big enough, and so, from
condition (ii) of Proposition \ref{equival}, the sequence of
integers $(\dim_\K{\rm ker}(\fJ_{k,i}^t))_{k\in \mathbb{N}_0}$ is
non-decreasing and becomes stationary for $k$ big enough.
\end{remark}

In the next section we proceed to study more closely this kind of
stationary properties related to the
rank of the matrices $\mathfrak{J}_{k,i}$.

\section{The rank of Jacobian sub-matrices}
\label{linearjacobian}

The aim of this section is to study the behavior of the ranks of
the matrices $\mathfrak{J}_{k,i}$ when $k$ and/or $i$ run over
$\N$ and $\mathbb{N}_{\ge e-1}$ respectively, which will provide
us with some information about certain invariants of the system
(\ref{sistema ampliado}), namely, the differentiation index and
the regularity of the Hilbert-Kolchin function.

We introduce a double sequence  $\mu_{k,i}$ of non-negative
integers associated with the matrices $\fJ_{k,i}$:

\begin{definition} \label{emes}
For $k\in \N_0$ and $i\in \mathbb{N}_{\ge e-1}$, we define
$\mu_{k, i}\in \N_0$ as follows:
\begin{itemize}
\item[--] $\mu_{0,i} := 0 $, for every $i\in \N_{\ge e-1}$;
 \item[--]
$\mu_{k,i}:=\dim _{\mathbb{K}}\ker (\mathfrak{J}_{k,i}^t)=k(n+r)-{\rm
rank}_{\mathbb{K}} (\mathfrak{J}_{k,i})$, for $k\ge 1$ and $i\in \N_0$.
\end{itemize}
\end{definition}

Now, we will focus on the study of certain stationarity properties
of the sequence $(\mu_{k,i})_{k,i}$. We begin by analyzing the
behavior of the sequence $(\mu_{k,i})_{k}$ when the index $i$ is
fixed, which will be done by comparing the matrices $\fJ_{k, i}$
for $k\in \N$.

First, let us observe the following obvious recursive relation which
holds for every $k\ge 1$:
\begin{equation}\label{remark2}
\mathfrak{J}_{k+1,i}=\left(
\begin{tabular}
[c]{ccccc}\cline{1-3}%
\multicolumn{1}{|c}{} &  &  & \multicolumn{1}{|c}{$0$} & $0$\\
\multicolumn{1}{|c}{} & $\mathfrak{J}_{k,i}\qquad$ &  &
\multicolumn{1}{|c}{$0$} & $0$\\
\multicolumn{1}{|c}{} &  &  & \multicolumn{1}{|c}{$\vdots$} &
$\vdots $\\\cline{1-3}
&  &  & 0 & 0\\
$\frac{\partial F^{(i-e+k+1)}}{\partial{X}^{(i+1)}}$ & $\frac{\partial F^{(i-e+k+1)}%
}{\partial U^{(i+1)}}$ & $\cdots$ & $\frac{\partial
F^{(i-e+k+1)}}{\partial
X^{(i+k+1)}}$ & $\frac{\partial F^{(i-e+k+1)}}{\partial U^{(i+k+1)}}$\\
$\frac{\partial G^{(i-e+k+1)}}{\partial{X}^{(i+1)}}$ & $\frac{\partial G^{(i-e+k+1)}%
}{\partial U^{(i+1)}}$ & $\cdots$ & $\frac{\partial
G^{(i-e+k+1)}}{\partial
X^{(i+k+1)}}$ & $\frac{\partial G^{(i-e+k+1)}}{\partial U^{(i+k+1)}}$%
\end{tabular}
\ \right)  .
\end{equation}
\bigskip

When the differential system (\ref{sistema ampliado}) is linear
(for instance, if $k:=\mathbb{Q}$ and the system (\ref{sistema
ampliado}) is of type $AU+B\dot{U}=Y$, where $A$ and $B$ belong to
$\mathbb{Q}^{r\times m}$), the matrices $\fJ_{k,i}$ have a nice
Hankel-block type form, but this is not exactly our situation.
However, there is a main relation arising from their underlying
differential structure which enables us to study them also in our
non linear setting:

\begin{proposition} \label{relacionderivadas}
Let $Z:=Z_1,\ldots ,Z_\alpha$ be differential independent
variables and let $H$ be a differential polynomial in $k\{Z\}$.
For all $l,j\in \mathbb{N}_0$ the following relation holds in
$\mathbb{K}$:
\begin{eqnarray}
\left(\dfrac{\partial H^{(l)}}{\partial Z^{(j+1)}}
\right)^{\centerdot} &=& \dfrac{\partial H^{(l+1)}}{\partial
Z^{(j+1)}} - \dfrac{\partial H^{(l)}}{\partial Z^{(j)}}.
\label{deriv1}
\end{eqnarray}
In particular, if $H\in \{F,G\}$ and $Z\in \{X,U\}$, we have that
$\dfrac{\partial H^{(l)}}{\partial Z^{(j+1)}}=0$ for every $j\ge
l+e$, since the order of $H^{(l)}$ is at most $l+e$, and
therefore identity (\ref{deriv1}) implies that
\begin{equation} \label{deriv2} \dfrac{\partial
H^{(l+1)}}{\partial Z^{(j+1)}}=\dfrac{\partial H^{(l)}}{\partial
Z^{(j)}} \qquad \forall\, j\ge l+e.
\end{equation}
Note that these identities are also valid over the differential
field $\K$, where the derivation is now the one induced by $\dot
X_j = f_j(X, U)$ in $\K$ and all partial derivatives are regarded
as elements in $\K$.

Due to the triangular form of $\mathfrak{J}_{k,i}$ and condition
(\ref{deriv2}), all the matrices $\mathfrak{J}_{k,i}$ have the same
$(n+r)\times (n+m)-$block $\mathfrak{J}_{1,i}$ in their main
diagonal.
\end{proposition}
\begin{proof}
Straightforward from the definitions and the Chain Rule.
\end{proof}

\bigskip

We are now ready to prove the first stationarity property of the
sequence $(\mu_{k,i})_{k,i}$ (see also Remark \ref{noether} above):

\begin{proposition}
\label{primero} For each fixed $i\in \mathbb{N}_{\ge e-1}$, the
sequence $(\mu_{k,i})_{k\in \N_0}$ is non-decreasing and verifies
the inequality \begin{equation} \label{cotas}
n\min\{k,e-1\}+\sum_{j=1}^{r} \min\{k,e-e_j\} \le \mu_{k,i} \le
\min \{k,e\}\, (n+r).
\end{equation}
In particular, there exists $k\in \N_0$
(depending on $i$), $0\le k\le e+n+\sum_{j=1}^{r} e_j$, such that
$\mu_{k,i}=\mu_{k+1,i}$.
\end{proposition}

\begin{proof}
The fact that $(\mu_{k,i})_{k}$ is a non-decreasing sequence
follows immediately by observing that $\ker(\fJ_{k,i}^t)\times \{
0 \} \subset \ker(\fJ_{k+1, i}^t)$ for every $k\in \N$ (see
(\ref{remark2}), also Remark \ref{noether}).

For every non-negative integer $k\in \N_0$, the matrix $\fJ_{k,i}$
has  $k(n+r)$ rows. Therefore, $\dim {\rm ker} (\fJ_{k,i}^t) \le
k(n+r)$. On the other hand, due to Proposition \ref{equival}, Remark
\ref{k0} and the definition of $\mu_{k,i}$, we have that ${\rm
trdeg}_\L ({\rm Frac}(A_i/\Delta_{i-e+1+k}\cap A_i)) = (m-r)(i+1) +
e(n+r) - \mu_{k,i}$. Now, ${\rm trdeg}_\L ({\rm
Frac}(A_i/\Delta_{i-e+1+k}\cap A_i))\ge {\rm trdeg}_\L ({\rm
Frac}(A_i/\Delta\cap A_i))$, since $\Delta_{i-e+1+k} \cap A_i
\subset \Delta \cap A_i$, and so, the fact that the differential
dimension of $\Delta$ is $m-r$ implies that ${\rm trdeg}_\L ({\rm
Frac}(A_i/\Delta\cap A_i)) \ge (m-r)(i+1) $. Hence, $\mu_{k,i} \le
e(n+r)$ holds.

In order to show the other inequality, we observe that, since the
order of the polynomial $G_j^{(p)}$ is $p+e_j$ $(1\le j \le r)$,
the partial derivatives ${\partial G_j^{(p)}}/{\partial X^{(q)}}$
and ${\partial G_j^{(p)}}/{\partial U^{(q)}}$ are all zero for
$q>p+e_j$. In particular, we conclude that for $t,s\in
\mathbb{N}$, with $t\ge s$, the derivatives ${\partial
G_j^{(i-e+t)}}/{\partial X^{(i+s)}}$ and ${\partial
G_j^{(i-e+t)}}/{\partial U^{(i+s)}}$ are zero if $i+s>i-e+t+e_j$,
or equivalently, if $e-e_j>t-s$. So, each polynomial $G_j$ induces
$\min\{k,e-e_j\}$ many null rows in the matrix
$\mathfrak{J}_{k,i}$. Analogously, each polynomial $F_j$ induces
$\min\{k,e-1\}$ many null rows in this matrix.

We conclude that the matrix $\mathfrak{J}_{k,i}$ has at least
$n\min\{k,e-1\}+\sum_{j=1}^{r} \min\{k,e-e_j\}$ null rows. Thus,
the dimension of the kernel of the transpose matrix
$\mathfrak{J}^t_{k,i}$ (i.e. $\mu_{k,i}$) is at least
$n\min\{k,e-1\}+\sum_{j=1}^{r} \min\{k,e-e_j\}$.

The second assertion follows directly from the fact that for every
$k\ge e$, the inequality (\ref{cotas}) reads
$n(e-1)+\sum_{j=1}^{r} (e-e_j)\le \mu_{k,i} \le e(n+r).$
\end{proof}

\bigskip

In fact, in Theorem \ref{segundo} we are able to prove a more
precise result than that of Proposition \ref{primero}: the sequence
$(\mu_{k,i})_{k\in \mathbb{N}_0}$ is strictly increasing up to a
certain index $k_i \le e+n+\sum_{j=1}^{r} e_j$ where it becomes
stationary.

For the sake of simplicity, we will use the following notations:

\begin{notation} \label{Hs}
The variables $X,U$ involved in system (\ref{sistema ampliado}) are
renamed in the following way: $Z_j:= X_j$ for $j=1,\ldots ,n$, and
$Z_{n+j}:=U_j$ for $j=1,\ldots ,m$ (and the same is done for their
corresponding formal derivatives). Analogously, the polynomials are
renamed as: $H_j:=F_j$ for $j=1,\ldots ,n$, and $H_{n+j}:=G_j$ for
$j=1,\ldots ,r$.
\end{notation}

With these notations, the matrix $\mathfrak{J}_{k,i}$ involves
exactly the derivatives of the polynomials $H^{(i-e+1+p)}$ with
respect to the variables $Z^{(i+q)}$, with $p=0,\ldots ,k-1$ and
$q=1,\ldots,k$.

\begin{theorem}
\label{segundo} Fix an index $i\in \N_{\ge e-1}$ and let $k_i\in
\N_0$ be the minimum of all the $k$'s in $\N_0$ such that
$\mu_{{k+1},i}=\mu_{{k},i}$ (this minimum is well defined due to
Proposition \ref{primero}). Then $\mu_{k,i}=\mu_{{k_i},i}$ for
every $k\ge k_i$.
\end{theorem}

\begin{proof} According to Notation \ref{Hs} we will rename
variables and equations as $Z:=(X,U)$ and $H:=(F,G)$.

The result is clear for $k_i = 0$: in this case, $\mu_{1,i} = 0$,
which is equivalent to the fact that the matrix
$\fJ_{1,i}=\dfrac{\partial H^{(i-e+1)}}{\partial {Z}^{(i+1)}} $
has full row rank. {}From identity (\ref{deriv2}) in Proposition
\ref{relacionderivadas} we conclude that $\mathfrak{J}_{k,i}$ has
full row rank too or, equivalently, that $\mu_{k,i}=0$ for all
$k$.

Now, let us assume that $k_i\ge 1$. It suffices to show that the
equality $\mu_{k,i}=\mu_{k-1 ,i}$ for an arbitrary index $k\ge 2$,
implies $\mu_{k+1,i}=\mu_{k,i}$.

In the sequel, for a vector $v\in \K^{l(n+r) }$ we will write its
description as a block vector $v = (v_1, \dots, v_{l})$ with $v_j
\in \K^{n+r}$.

Due to the recursive relation (\ref{remark2}), the identity
${\ker}(\mathfrak{J}_{k,i}^t)\times \{0\}=
{\ker}(\mathfrak{J}_{k+1,i}^t)\cap \{ v_{k+1} = 0\}$ holds in
$\mathbb{K}^{(k+1)(n+r)}$ for every $k\in \N$ and so, the
equality $\mu_{k, i} = \mu_{k+1,i}$ is equivalent to the
inclusion ${\ker}(\mathfrak{J}_{k+1,i}^t) \subset \{ v_{k+1} =
0\}$. Then, the theorem is a consequence of the following
recursive principle:

\vskip8pt

\noindent \textit{Claim:} \textit{For all $k\in \N$,
${\ker}(\mathfrak{J}^t_{k,i}) \subset \{ v_{k} = 0\}$ implies
${\ker}(\mathfrak{J}^t_{k+1,i}) \subset \{ v_{k+1} = 0\}$}.

\vskip4pt

\noindent \textit{Proof of the Claim.- } Let us show that if
$(v_1, \dots, v_{k+1}) \in \ker(\mathfrak{J}^t_{k+1,i})$ then, the
vector $w  = (w_1,\dots, w_k)\in \K^{k (n+r)}$ defined as
$$
w_{k} := v_{k+1}, \qquad w_j := v_{j+1} - \dot w_{j+1}, \quad  j =
k-1, \dots, 1,$$ lies in $\ker (\mathfrak{J}^t_{k,i})$, which
implies the Claim.

Our assumptions on the order of the equations imply that
$\dfrac{\partial H^{(i-e+\ell)}}{\partial Z^{(i+j)}} = 0$ for $\ell
<j$. Hence, we have that $w \in \ker (\mathfrak{J}^t_{k,i})$ if and
only if the following identities hold over $\K$ for every $1\le j
\le k$:
$$\displaystyle\sum_{\ell=j}^{k} w_{\ell} \ \dfrac{\partial
H^{(i-e+\ell)}}{\partial Z^{(i+j)}}= 0.$$

We will proceed recursively for $j= k, k-1, \dots, 1$. For $j=k$,
the definition of $w$ and identity (\ref{deriv2}) imply that
\[w_{k}\ \dfrac{\partial H^{(i-e+k)}}{\partial Z^{(i+k)}} =
v_{k+1}\ \dfrac{\partial H^{(i-e+k+1)}}{\partial Z^{(i+k+1)}} =
0.\]

Now, assume that $\displaystyle\sum_{\ell=j+1}^{k} w_{\ell} \
\dfrac{\partial H^{(i-e+\ell)}}{\partial Z^{(i+j+1)}}= 0$.
Differentiating this identity in $\K$ and using identity
(\ref{deriv1}) we get:
$$\sum_{\ell=j+1}^{k} \dot
w_{\ell} \ \dfrac{\partial H^{(i-e+\ell)}}{\partial
Z^{(i+j+1)}}+\sum_{\ell=j+1}^{k}  w_{\ell} \ \left(\dfrac{\partial
H^{(i-e+\ell+1)}}{\partial Z^{(i+j+1)}} - \dfrac{\partial
H^{(i-e+\ell)}}{\partial Z^{(i+j)}}\right) = 0.$$ This implies that
\begin{eqnarray*}
\sum_{\ell = j}^{k} w_{\ell} \ \dfrac{\partial
H^{(i-e+\ell)}}{\partial Z^{(i+j)}} &=& w_{j} \ \dfrac{\partial
H^{(i-e+j)}}{\partial Z^{(i+j)}} + \sum_{\ell=j+1}^{k} \dot w_{\ell}
\ \dfrac{\partial H^{(i-e+\ell)}}{\partial
Z^{(i+j+1)}}+\sum_{\ell=j+1}^{k} w_{\ell} \ \dfrac{\partial
H^{(i-e+\ell+1)}}{\partial Z^{(i+j+1)}}\\
&=& \sum_{\ell=j+1}^{k} (\dot w_{\ell} +w_{\ell - 1}) \
\dfrac{\partial H^{(i-e+\ell)}}{\partial Z^{(i+j+1)}} +w_k\
\dfrac{\partial
H^{(i-e+k+1)}}{\partial Z^{(i+j+1)}}\\
&=& \sum_{\ell=j+1}^{k+1} v_\ell \ \dfrac{\partial
H^{(i-e+\ell)}}{\partial Z^{(i+j+1)}} \ = \ \sum_{\ell=1}^{k+1}
v_\ell \ \dfrac{\partial H^{(i-e+\ell)}}{\partial Z^{(i+j+1)}} \ = \
0,
\end{eqnarray*}
where the second equality follows from identity (\ref{deriv2})
and the third one is simply the definition of  $w$. This concludes
the proof of the theorem.
\end{proof}

\bigskip

So far, we have studied the behavior of the sequence $(\mu_{k,i})_k$
for an arbitrary (but fixed) index $i\in \mathbb{N}_{\ge e-1}$. In
the remaining part of the section we will analyze the sequence
$(\mu_{k,i})_i$ fixing the index $k\in \N$.

We start by exhibiting a (non $\mathbb{K}$-linear) bijection between
the kernels of the matrices $\mathfrak{J}_{k,i}$ and
$\mathfrak{J}_{k,i+1}$ for any index $k\in \N$.

\begin{lemma} \label{ii1}
Let $(v_1,\ldots, v_{k})$ and $(w_1,\ldots ,w_{k})$ be arbitrary
elements in $\mathbb{K}^{k(n+m)}$ (here $v_j$ and $w_j$ denote
vectors in $\mathbb{K}^{n+m}$ for every index $j$). Then, for each
$i\in \mathbb{N}_{\ge e-1}$, the function
$\theta:\mathbb{K}^{k(n+m)} \to \mathbb{K}^{k(n+m)}$ defined as $
\theta(v_1,\ldots, v_{k})= (v_1,v_2-\dot{v}_1,v_3-\dot{v}_2,\ldots,
v_{k}-\dot{v}_{k-1}) $ maps $\ker (\mathfrak{J}_{k,i+1})$ into $\ker
(\mathfrak{J}_{k,i})$. Moreover, $\theta$ is a bijection between
$\ker (\mathfrak{J}_{k,i+1})$ and $\ker (\mathfrak{J}_{k,i})$, with
inverse $ \theta^{-1}(w_1,\ldots
,w_{k})=(w_1,{w_2}+\dot{w}_1,{w_3}+\dot{w}_2+w_1^{(2)},\ldots
,w_{k}+\dot{w}_{k-1}+\cdots +w_1^{(k-1)}). $
\end{lemma}

\begin{proof} It is easy to see that $\theta$ is a bijection in
$\K^{k(n+m)}$ with the inverse given in the statement of the
Lemma. Let us show that it maps $\ker (\mathfrak{J}_{k,i+1})$ to
$\ker (\mathfrak{J}_{k,i})$.

We keep the notations introduced in Notation \ref{Hs}.

For arbitrary vectors $(v_1,\dots, v_k)$ and $(w_1,\dots, w_k)$ in
$\K^{k(n+m)}$, consider the following two families of sums
$(p=0,\dots, k-1)$:
$$E_{i+1,p}(v) := \sum_{j=1}^{k} \dfrac{\partial H^{(i-e+2+p)}}{\partial
Z^{(i+1+j)}}\ v_j \quad \hbox{ and } \quad  E_{i,p}(w)
:=\sum_{j=1}^{k} \dfrac{\partial H^{(i-e+1+p)}}{\partial
Z^{(i+j)}}\ w_j.$$ Note that $v \in \ker(\fJ_{k, i+1})$ if and
only if $E_{i+1,p}(v) = 0$ for $p=0, \dots, k-1$ and $w \in
\ker(\fJ_{k, i})$ if and only if $E_{i,p}(w) = 0$ for $p=0,\dots,
k-1$.

First, we compare the vectors $\fJ_{k, i+1}\, v$ and $\fJ_{k, i}\,
\theta(v)$ for a given vector $v=(v_1,\dots, v_k)\in \K^{k(n+m)}$:
let us observe that $E_{i+1,0}(v) = E_{i,0}(v) =
E_{i,0}(\theta(v))$, since $\dfrac{\partial H^{(i-e+2)}}{\partial
Z^{(i+1+j)}} = \dfrac{\partial H^{(i-e+1)}}{\partial Z^{(i+j)}} =
0$ for every $j\ge 2$ (recall that the order of $H$ is smaller
than $e$), $\dfrac{\partial H^{(i-e+2)}}{\partial Z^{(i+2)}} =
\dfrac{\partial H^{(i-e+1)}}{\partial Z^{(i+1)}}$ (from
(\ref{deriv2})) and $\theta(v)_1 = v_1$. Now, for $p>0$, we have:
$$\begin{array}{ccc}{E}_{i+1,p}(v) - (E_{i+1,p-1}(v))^{\centerdot}
&=& \hspace{8cm} \\
\end{array}
$$
\begin{eqnarray*}
&=&\sum_{j=1}^{k} \left( \dfrac{\partial H^{(i-e+2+p)}} {\partial
Z^{(i+1+j)}}-\left(\dfrac{\partial H^{(i-e+1+p)}}{\partial
Z^{(i+1+j)}}\right)^{\centerdot}  \right) v_j - \sum_{j=1}^{k}
\dfrac{\partial H^{(i-e+1+p)}}{\partial Z^{(i+1+j)}}\ \dot{v_j}\\
&=&\sum_{j=1}^{k} \dfrac{\partial H^{(i-e+1+p)}}{\partial Z^{(i+j)}}
v_j - \sum_{j=2}^{k+1} \dfrac{\partial H^{(i-e+1+p)}}{\partial
Z^{(i+j)}}\ \dot{v}_{j-1} \\
&=& \dfrac{\partial H^{(i-e+1+p)}}{\partial Z^{(i+1)}} v_1
+\sum_{j=2}^{k} \dfrac{\partial H^{(i-e+1+p)}}{\partial Z^{(i+j)}}
(v_j - \dot{v}_{j-1}) \ = \ E_{i,p}(\theta(v)),
\end{eqnarray*}
where the second equality follows from identity (\ref{deriv1}),
and the third one from the fact that $\dfrac{\partial
H^{(i-e+1+p)}}{\partial Z^{(i+k+1)}} = 0$ for $p\le k-1$.

These equalities imply straightforwardly that $\theta$ maps
$\ker(\fJ_{k, i+1})$ into $\ker(\fJ_{k, i})$.

In order to prove that it is onto, we may argue recursively: if $w
\in \ker(\fJ_{k, i})$, then $E_{i,p}(w) = 0$ for $p=0, \dots,
k-1$. Now, $E_{i+1,0}(\theta^{-1}(w)) = E_{i+1,0}(w) = E_{i,0}(w)
= 0$. Assuming that $E_{i+1,p-1}(\theta^{-1}(w)) = 0$ has already
been proved, we deduce that $E_{i+1,p}(\theta^{-1}(w))  =
E_{i,p}(w) + (E_{i+1,p-1}(\theta^{-1}(w))^\centerdot $ also equals
$0$. We conclude that $\theta^{-1}(w) \in \ker(\fJ_{k, i+1})$.
\end{proof}

\bigskip

Even though the bijection $\theta$ between $\ker(\fJ_{k, i+1})$
and $\ker(\fJ_{k, i})$ shown in the previous Lemma is not a
$\K$-linear map, it enables us to prove the following:

\begin{proposition} \label{rangocorrido}
Let $k\in \mathbb{N}_0$ and $i\in \mathbb{N}_{\ge e-1}$ be arbitrary
integers. Then $\mu_{k,i}=\mu_{k,i+1}$.
\end{proposition}

\begin{proof}
The result is immediate if $k=0$. Now, let $k\in \mathbb{N}$ be a
positive integer. In order to prove that $\dim_\K(\ker(\fJ_{k,
i}^t) ) = \dim_\K(\ker(\fJ_{k, i+1}^t) )$, it suffices to show
that $\mathfrak{J}_{k,i}$ and $\mathfrak{J}_{k, i+1}$ have the
same rank, since they are two matrices of the same size.

For each pair of indices $j,t$, $1\le j \le k$ and $1\le t\le n+m$,
set $$C_{j,t}:=\dfrac{\partial H^{[i-e+1,i-e+k]}}{\partial
Z_t^{(i+j)}} \quad \hbox{ and } \quad D_{j,t}:=\dfrac{\partial
H^{[i-e+2,i-e+k+1]}}{\partial Z_t^{(i+1+j)}}$$ for the corresponding
columns of the matrices $\mathfrak{J}_{k,i}$ and
$\mathfrak{J}_{k,i+1}$, respectively.

Assume that a column $D_{j_0,t_0}$ of the matrix
$\mathfrak{J}_{k,i+1}$ is a $\mathbb{K}$-linear combination of
the columns $D_{j,t}$ to its right. Then, there exist elements
$\alpha_{j,t}\in \mathbb{K}$ such that
\[
D_{j_0,t_0}=\sum_{t=t_0+1}^{n+m} \alpha_{j_0,t}D_{j_0,t} +
\sum_{j=j_0+1}^{k} \sum_{t=1}^{n+m}\alpha_{j,t}D_{j,t},
\]
and so, the vector $
v:=(\overrightarrow{0},\ldots,\overrightarrow{0},-\alpha_{j_0},
-\alpha_{j_0+1},\ldots,-\alpha_{k})\in \mathbb{K}^{k(n+m)} $
belongs to the kernel of $\mathfrak{J}_{k,i+1}$, where
$\overrightarrow{0}$ denotes the null vector in
$\mathbb{K}^{n+m}$, $\alpha_{j_0}:=(0,\dots, 0, 1, \alpha_{j_0,
t_0+1}, \dots, \alpha_{j_0, k})$ and $\alpha_{j}:=(\alpha_{j,1},
\dots, \alpha_{j, k})$ for $j\ge j_0+1$. By Lemma \ref{ii1}, the
vector $\theta(v)$ belongs to the kernel of $\mathfrak{J}_{k,i}$
and, due to the particular form of the application $\theta$, it
turns out that the column $C_{j_0, t_0}$ is a $\K$-linear
combination of the columns to its right. Hence, the rank of
$\mathfrak{J}_{k,i}$ is lower than or equal to the rank of
$\mathfrak{J}_{k,i+1}$.

By means of the inverse map $\theta^{-1}$, one proves in an
analogous way that the rank of $\mathfrak{J}_{k,i+1}$ is lower
than or equal to the rank of $\mathfrak{J}_{k,i}$ .
\end{proof}

\begin{remark}
The following alternative proof of this proposition was kindly suggested to us
by Prof. F. Ollivier. From \cite[Prop.~10, Ch.~IV]{kol} the dimensions of the
kernels of  $\mathfrak{J}_{k,i}$ and $\mathfrak{J}_{k,i+1}$ are the
differential dimensions of the ideals generated in a differential polynomial
ring in $k(n+m)$ new variables by the linear equations defined by these
matrices. Now, Proposition \ref{rangocorrido} follows since the bijection
$\theta$ induces a differential isomorphism between these fields.
\end{remark}

The previous proposition states that the sequence $\mu_{k,i}$ does not depend
on the index $i$; therefore:

\begin{notation} \label{sacar i}
In the sequel, we will write $\mu_k$ instead of $\mu_{k,i}$, for
any $i\in \mathbb{N}_{\ge e-1}$.
\end{notation}

So, Theorem \ref{segundo} can be restated as follows:

\begin{corollary} \label{ki constante}
There exists a non-negative integer $\sigma$ such that
$\mu_k<\mu_{k+1}$ for every $k<\sigma$ and $\mu_k=\mu_{k+1}$ for
every $k\ge \sigma$. \square
\end{corollary}

\section{The differentiation index} \label{seccind}

Here we apply the properties of the matrices $\mathfrak{J}_{k,i}$
established in the previous section to the study of well-known
invariants of our DAE system (\ref{sistema ampliado}): the
differentiation index, the regularity of the Hilbert-Kolchin
function and the order of its associated differential ideal.

We start by introducing the notion of differentiation index of the
DAE system (\ref{sistema ampliado}) (see also \cite{brenan},
\cite{Fliess}, \cite{campbell-gear}, \cite{levey}):

\begin{definition} \label{indice}
The non negative integer $\sigma$ introduced in Corollary \ref{ki constante}
is called the \emph{differentiation index}
of the system (\ref{sistema ampliado}).
\end{definition}

Corollary \ref{ki constante} states that the differentiation
index $\sigma$ is the smallest non-negative integer $k$ where the
sequence $\mu_{k}$ becomes stationary. Since $\mu_\sigma=\mu_k$
for all $k\ge \sigma$, we deduce from Proposition \ref{primero}:

\begin{remark} \label{inequalities} The following inequalities hold:
\begin{equation} \label{Bezout0}
(e-1)n+\sum_{j=1}^{r} (e-e_j)\le \mu_\sigma \le e(n+r);
\end{equation}
\begin{equation} \label{remark indice}
0\le \sigma\le \min\bigg\{ e(n+r)\ ,\ e+n+\sum_{j=1}^{r}
e_j\bigg\}.
\end{equation}
\end{remark}

For the last inequality see also \cite[Section 5.2]{sadikb}.

\subsection{The manifold of constraints}
\label{sectionindex}

For every $i\in \N_{\ge e-1}$, the differentiation index $\sigma$ is strongly
related to the minimum number of derivatives of the system (\ref{sistema
ampliado}) required to obtain the intersection of the whole differential ideal
$\Delta$ with the polynomial ring $A_i$, namely those polynomials in $\Delta$
which involve only derivatives up to order $i$ (a similar result can also be
obtained by rewriting techniques by means of \cite[Theorem 27] {sadikb}; see
also \cite[Lemma 14]{djs} and \cite[Lemma 9]{MS}):

\begin{theorem} \label{sadik}
Let $\sigma \in \N_{0}$ be the differentiation index of the system
(\ref{sistema ampliado}). Then, for every $i\in \N_{\ge e-1}$, the
equality of ideals $$\Delta_{i-e+1+\sigma}\cap A_i=\Delta \cap
A_i$$ holds in the polynomial ring $A_i$. In particular, from
inequality (\ref{remark indice}) it follows that the identity
$\Delta_{i+1+n+\sum_{j=1}^{r} e_j}\cap A_i = \Delta \cap A_i$
holds for every $i\in \N_{\ge e-1}$. Furthermore, for every $i\in
\N_{\ge e-1}$, the differentiation index $\sigma$ verifies:
$\sigma= \min\{ h\in \N_0 : \Delta_{i-e+1+h} \cap A_i = \Delta
\cap A_i\}.$ Taking $i=e-1$, we have
\begin{equation}\label{indexmin}
\sigma= \min\{ h\in \N_0 : \Delta_{h} \cap A_{e-1} = \Delta \cap
A_{e-1}\}.
\end{equation}
\end{theorem}

\begin{proof} Fix an index $i\in \N_{\ge e-1}$.
Let us consider the increasing chain $(\Delta_{i-e+1+k}\cap
A_i)_{k\in \N_0}$ of prime ideals in the polynomial ring $A_i$.
{}From Proposition \ref{equival}, Remark \ref{k0} and the definition
of the sequence $\mu_k$, for every $k\in \N_0$, we have
\begin{equation}\label{trdegmu}
{\rm trdeg}_\L ({\rm Frac}(A_i/\Delta_{i-e+1+k}\cap A_i))
 =  (m-r)(i+1) + e(n+r) - \mu_{k}.
\end{equation}
Since $\mu_{k}$ is stationary for $k\ge \sigma$ (Corollary
\ref{ki constante}), the previous equality implies that all the
prime ideals $\Delta_{i-e+1+k}\cap A_i$ have the same dimension
for $k\ge \sigma$, and so, they coincide because they form an
increasing chain of ideals. On the other hand, any finite system
of generators of the prime ideal $\Delta\cap A_i\subset A_i$
belongs to $\Delta_{i-e+1+k}\cap A_i$ for all $k$ big enough,
which finishes the proof of the first assertion of the Theorem.

In order to prove the second part of the statement, for each $i\in
\mathbb{N}_{\ge e-1}$, let $h_i$ be the smallest non-negative
integer such that $\Delta_{i-e+1+h_i}\cap A_i  = \Delta\cap A_i $.
By the definition of $h_i$, the transcendence degrees ${\rm
trdeg}_\L ({\rm Frac}(A_i/\Delta_{i-e+1+k}\cap A_i))$ coincide for
$k\ge h_i$, and so, $\mu_{k} $ is constant for $k\ge h_i$ (see
identity (\ref{trdegmu}) above). This implies that $\sigma  \le
h_i$. The equality follows from the first part of the statement and
the minimality of $h_i$.
\end{proof}

\bigskip

Identity (\ref{indexmin}) in Theorem \ref{sadik} can be regarded as
an alternative definition of the differentiation index (see also
\cite[Section 3.2]{MS} for first-order DAE systems). In particular,
it gives the following interpretation of the differentiation index
$\sigma$ (see \cite{Fliess} for the case $e=1$):

\begin{remark}
If the differentiation index $\sigma$ of system (\ref{sistema
ampliado}) equals $0$, there are no constraints on initial
conditions for the system (recall that $\Delta_0:=(0)$). In the
case when $\sigma \ge 1$, the quantity $\sigma-1$ is the minimal
number of derivatives of the equations in the system needed to
obtain all the relations that the initial conditions must satisfy
(the so called ``manifold of constraints on initial conditions").
\end{remark}

Another fundamental property of the differentiation index,
concerning the number of derivatives  of the equations required to
obtain an explicit equivalent ODE system, will be considered in
Subsection \ref{ODEdim0} and Section \ref{ODE} below.

First we will show how our previous results can be applied in
order to estimate the regularity of the Hilbert-Kolchin function.

\subsection{The regularity of the Hilbert-Kolchin function and applications}
\label{Hilbertfunction}

We recall the main basic facts of the Hilbert-Kolchin function
for the particular case of our input system (\ref{sistema
ampliado}) and our ideal $\Delta$ (see \cite[Chapter II]{kol} for
the general theory).

The \emph{Hilbert-Kolchin function}
$H_\Delta:\mathbb{N}_0\rightarrow \mathbb{N}_0$ of the differential
ideal $\Delta\subset \mathbb{L}\{X,U\}$ is defined as
$$H_\Delta(i):=\textrm{trdeg}_\L(\textrm{Frac}(A_i/(\Delta \cap
A_i)))$$ for every $i\in \N_0$. Since the ideal $\Delta$ we are
considering has differential dimension $m-r$, the identity
$H_\Delta(i)  = (m-r)(i+1) + {\rm ord}_\L(\Delta)$ holds for $i$
\emph{sufficiently big} (see for instance \cite[Ch.~II, Sec.~12,
Th.~6]{kol}), where ${\rm ord}_\L(\Delta)$ is a non-negative integer
depending only on the differential ideal $\Delta$ which is called
the \emph{order} of the ideal. The polynomial
$$\mathcal{H}_\Delta (T):=(m-r)(T+1)+ {\rm ord}_\L(\Delta)$$ is
called the \emph{Hilbert-Kolchin polynomial} associated to the
ideal $\Delta$. The \emph{regularity of the Hilbert-Kolchin
function} $H_\Delta$ is defined to be the minimum integer $i_0\in
\mathbb{N}_0$ such that $H_\Delta(i)=\mathcal{H}_\Delta(i)$ holds
for all $i\ge i_0$.

\subsubsection{The regularity of the Hilbert-Kolchin
function}\label{secregularity}

The theory of \emph{characteristic sets} (see for instance
\cite{ritt}) can be used to give a precise estimation of the
regularity of $H_\Delta$: let $\mathcal{C}$ be a characteristic set
of the differential ideal $\Delta \subset \mathbb{L}\{X,U\}$ for an
\emph{orderly ranking} (see \cite{ritt}) in the variables $X,U$.
Then, the regularity of $H_\Delta$ is equal to $\max \{\text{ord}(C)
: C \in \mathcal{C}\}-1$ (this fact follows as an immediate
consequence of the proof of \cite[Ch.~II, Section 12, Th.~6
(d)]{kol}, see also \cite[Theorem 3.3]{carra}).

The results developed so far enable us to exhibit the following
simple upper bound for this regularity,  which depends only on the
order of the polynomials involved in the system (\ref{sistema
ampliado}):

\begin{theorem} \label{hilbert constante}
The regularity of the Hilbert-Kolchin function of the ideal
$\Delta$ over the ground field $\mathbb{L}$ is bounded by $e-1$.

In particular, for first-order systems of type (\ref{sistema
ampliado}) (in other words, for the case $e=1$), the Hilbert-Kolchin
function of $\Delta$ coincides with the associated Hilbert-Kolchin
polynomial.
\end{theorem}

\begin{proof} It suffices to show that the equality $H_\Delta(i+1) = H_\Delta(i) + (m-r)$
holds for every $i\in \N_{\ge {e-1}}$.

Fix an index $i\in \mathbb{N}_{\ge e-1}$. Due to Theorem
\ref{sadik}, we have that $\Delta \cap A_i = \Delta_{i+1-e+\sigma}
\cap A_i$ and $\Delta \cap A_{i+1} = \Delta_{i+2-e+\sigma} \cap
A_{i+1}$. So, $H_\Delta(i+1) = {\rm trdeg}_\L(A_{i+1}
/(\Delta_{i+2-e+\sigma} \cap A_{i+1}))$ and $H_\Delta(i) = {\rm
trdeg}_\L(A_i / (\Delta_{i+1-e+\sigma} \cap A_i))$. Thus, using
Proposition \ref{equival} (recall also Notation \ref{sacar i}) we
obtain:
\begin{eqnarray*}
H_\Delta(i+1) &=& (m-r)(i+2) + e(n+r) - \mu_{\sigma},\\
H_\Delta(i) &=& (m-r)(i+1) + e(n+r) - \mu_{\sigma}.
\end{eqnarray*}
Hence, the equality $H_\Delta(i+1) = H_\Delta(i) + (m-r)$ holds.
\end{proof}

\bigskip

As we have already pointed out, the regularity of the
Hilbert-Kolchin function can be obtained from the maximal order
appearing in any characteristic set associated to the differential
system (\ref{sistema ampliado}) for an orderly ranking. Therefore,
our upper bound on the regularity implies an upper bound for this
maximum:

\begin{corollary} Let $\mathcal{C}$ be a characteristic set of
$\Delta\subset \mathbb{L}\{X,U\}$ with respect to an orderly ranking
on the variables $X,U$. Then, the order of each of the differential
polynomials in $\mathcal{C}$ is bounded by the maximal order
$e:=\max \{ 1, e_1,e_2,\ldots ,e_r\}$.  \square
\end{corollary}

\subsubsection{Upper bounds for the order of the ideal $\Delta$} \label{orders}

The proof of Theorem \ref{hilbert constante} also shows an indirect
relation between the Hilbert-Kolchin polynomial $\mathcal{H}_\Delta$
and the differentiation index $\sigma$. More precisely:

\begin{remark} \label{indice-hilbert}
The Hilbert-Kolchin polynomial $\mathcal{H}_\Delta(T)$ of the ideal
$\Delta$ can be written as $\mathcal{H}_\Delta(T) = (m-r)(T+1) +
e(n+r) - \mu_\sigma$. Equivalently, ${\rm{ord}}_{\, \mathbb
L}(\Delta)=e(n+r)-\mu_\sigma$.
\end{remark}

{}From inequality (\ref{Bezout0}) and Remark \ref{indice-hilbert} we
deduce the following upper bound for the order of the ideal $\Delta$
in terms of the orders of the defining equations:

\begin{corollary} \label{ritt-jacobi} The order of the ideal
$\Delta$ can be bounded as follows:
 \[{\rm{ord}}_{\, \mathbb
L}(\Delta)\le n+\sum_{j=1}^r e_j. \ \square\]
\end{corollary}

This result is known as Greenspan's bound. It was first proved by B. Greenspan
in \cite{greenspan} for difference equations and extended later by R. Cohn in
\cite{cohn} for $0$-dimensional components of differential systems.

Another upper bound for the order is given by Ritt in \cite[Ch.~VII,
p.135]{ritt} (see also \cite[Chapter IV, Proposition 9]{kol} and \cite{sadikt})
and holds for general 0-dimensional DAE systems: the order of its associated
differential ideal is always bounded by the sum of the maxima of the orders of
each variable in the given differential polynomials. This bound is called by
Ritt a \emph{differential analogue to B\'ezout's Theorem}. In our case, we are
able to give a simple proof of the following extension of this fact:

\begin{corollary} \label{ritt-jacobi vero}
For each $i=1,\ldots m$ set $\epsilon_i:=\max \{ {\rm
ord}_{U_i}G_j,\ j=1,\ldots ,r\}$. We have the following inequality:
\[
{\rm{ord}}_{\, \mathbb L}(\Delta)\le n+\sum_{i=1}^m \epsilon_i.
\]
\end{corollary}

\begin{proof} We transform our input system (\ref{sistema ampliado}) in an equivalent
\emph{first-order} system by introducing new variables and equations
in the usual way: for each $i=1,\ldots, m$ and $\ell=0,\ldots,
\epsilon_i$ we define $U_{i,\ell}:=U_i^{(\ell)}$. Making this
substitution in the original system and adding the equations
$U_{i,\ell}=\dot{U}_{i,\ell-1}$ ($1\le i\le m$, $1\le \ell\le
\epsilon_i$) we obtain a first-order DAE system $(\Sigma')$ with a
structure similar to the original system (\ref{sistema ampliado}),
but such that the order of each parametric equation $g_j=Y_j$ is $0$
(in other words, the new $e_j$'s are zero for $j=1,\ldots ,r$).

Set $\mathcal{U}$ for the set of variables $U_{i,\ell}$ with $1\le
i\le m$, $0\le \ell\le \epsilon_i$. For each integer $p\ge 0$,
denote by $\Delta'\subset k\{X,\mathcal{U}\}$ the differential ideal
generated by the equations defining $(\Sigma')$ and by $A'_p$ the
polynomial ring $\mathbb{L}[X^{[p]},\mathcal{U}^{[p]}]$.

It is easy to see that for each $p>e$ the map
\[
U_i^{(j)}\mapsto \left\{
\begin{array}[c]{lcl}%
U_{i,j} & \text{if} & 0\le j\le \epsilon_i \\[1.5mm]
U_{i,\epsilon_i}^{(j-\epsilon_i)} &\text{if}& \epsilon_i\le j\le p \\
\end{array}
\right.
\]
induces a monomorphism $A_p/(\Delta\cap A_p)\hookrightarrow
A'_p/(\Delta'\cap A'_p)$. Hence, for $p$ sufficiently big, the
inequality $\mathcal{H}_{\Delta}(p)\le \mathcal{H}_{\Delta'}(p)$
holds. Since the differential dimension of both systems
(\ref{sistema ampliado}) and $(\Sigma')$ is $m-r$, we conclude that
$ \text{ord}_{\mathbb{L}}(\Delta)\le
\text{ord}_{\mathbb{L}}(\Delta'). $

The proof finishes by applying Corollary \ref{ritt-jacobi} to the
first-order system $(\Sigma')$, and observing that in this case the
bound is simply the number of ``non parametric" equations (since the
new parametric ones have order zero), which is exactly
$n+\sum_{i=1}^m \epsilon_i$.
\end{proof}

\bigskip
An interesting and more precise B\'ezout-type bound for $0$-dimensional systems
(which includes both Corollaries \ref{ritt-jacobi} and \ref{ritt-jacobi vero})
is discussed by Jacobi in \cite{jacobi} (see also \cite[Ch.VII, \S 6]{ritt}).
Even though Jacobi's bound remains conjectural for the general case, in
\cite{kondra} (based  on \cite{ritt2}) it has been proved  to hold  for a class
of $0$-dimensional differential ideals including those considered in this
paper. More precisely, suppose $m=r$ and set $e_{ij}$ for the maximum of the
orders of a derivative of the variable $U_i$ occurring in $g_j$, putting
$e_{ij}:=-\infty$ if the variable $U_i$ does not appear in $g_j$. Then,
Jacobi's inequality states that
\[
\textrm{ord}_{\mathbb{L}}(\Delta)\le n+\displaystyle{\max_{\tau\in
\mathcal{S}_m} \sum_{j=1}^{m}e_{\tau(j)j}},\] where $\mathcal{S}_m$
is the permutation group of order $m$. Now we show an example where
Jacobi's bound is reached and improves the previous ones.

\begin{example} Consider the following first-order DAE system
with coefficients in $k:=\mathbb{Q}$: $$  \left\{
\begin{array}
[c]{ccl}%
Y_1 & = & U_1\\
Y_2 & = & U_2+\dot{U}_1\\
 &\vdots  & \\
Y_{m} & = & U_{m}+\dot{U}_{m-1}
\end{array}
\right. $$ with $m>2$. Here, both upper bounds for ${\rm
ord}_{\mathbb{L}}(\Delta)$ given by Corollaries \ref{ritt-jacobi}
and \ref{ritt-jacobi vero} equal $m-1$, while Jacobi's upper bound
is $0$. It is not difficult to see that ${\rm
ord}_{\mathbb{L}}(\Delta) = 0$.
\end{example}

\subsubsection{The index and an equivalent explicit ODE system}
\label{ODEdim0}

The estimation for the regularity of the Hilbert-Kolchin function
allows us to give a first result concerning the number of
derivatives of the input equations required to obtain an equivalent
explicit ODE system in the $0$-dimensional case. We will show that
this number is at most the differentiation index of the system:

\begin{theorem} \label{despeje general dimension 0} Let $(\Sigma)$ be a DAE system
as in (\ref{sistema ampliado}) of differential dimension $0$ (or
equivalently, $r=m$), maximal order bounded by $e$ and
differentiation index $\sigma$. Let $\Xi =\{
\xi_1^{(\ell_1)},\ldots,\xi_s^{(\ell_s)}\}\subseteq \{
X^{[e-1]},U^{[e-1]}\}$ be an algebraic transcendence basis of the
fraction field of $A_{e-1}/\Delta\cap A_{e-1}$ over the field
$\mathbb{L}$. Then:
\begin{enumerate}

\item for each $i=1,\ldots ,s$ there exists a non-zero separable polynomial
$P_i$ with coefficients in the base field $k$, such that $P_i(Y^{[\sigma]},\Xi,
\xi_i^{(e)})\in (F^{[\sigma]}, G^{[\sigma]})\subset k[Y^{[\sigma]},
X^{[\sigma+1]}, U^{[\sigma + e]}]$;

\item  set $\{\eta_{s+1}, \ldots , \eta_{n+m}\}:=\{X, U\}\setminus \{\xi_1,
\ldots , \xi_s\}$. Then, 
for all $i=s+1,\ldots ,n+m$, there exists a non-zero separable polynomial $P_i$
with coefficients in the base field $k$, such that $P_i(Y^{[\sigma-1]},\Xi,
\eta_i^{(e-1)})\in (F^{[\sigma-1]}, G^{[\sigma-1]})\subset k[Y^{[\sigma-1]},
X^{[\sigma]}, U^{[\sigma+e-1]}]$.
\end{enumerate}

In particular, for every $\eta\in \{X,U\}$ there exists a separable
non-trivial polynomial relation between $\eta^{(e)}$ and $\Xi$
modulo $\Delta$ which can be obtained using at most $\sigma$
derivations of the input equations.
\end{theorem}

\begin{proof} In order to prove the first statement, we fix a variable
$\xi_{i_0}$. Since we are in a differential $0$-dimensional situation, from the
upper bound on the regularity of the Hilbert-Kolchin function (Theorem
\ref{hilbert constante}), we have that the set $\Xi$ is also an algebraic
transcendence basis of the fraction field of the ring $A_{e}/\Delta\cap A_{e}$.
Then, there exists a polynomial $Q$ in $s+1$ variables with coefficients in
$\mathbb{L}$, such that $Q(\Xi,\xi_{i_0}^{(e)})$ belongs to the ideal
$\Delta\cap A_e=\Delta_{\sigma+1}\cap A_e$ (Theorem \ref{sadik}). Clearly, this
polynomial can be chosen so that $\dfrac{\partial Q}{\partial
\xi_{i_0}^{(e)}}\notin \Delta$ (i.e. separable). Now, the polynomial $P_{i_0}$
of the statement can be easily obtained from the previous $Q$ by multiplying it
by an adequate factor in $k\{ Y \}$ and evaluating the superfluous variables
$Y^{(l)}$ for all $l\ge \sigma+1$ at suitably chosen elements of the base field
$k$.

The second assertion follows similarly, but in this case we use the fact that
the family  $\{\eta_i^{(e-1)}, \Xi\}$ is algebraically dependent when regarded
in the fraction field of $A_{e-1}/\Delta\cap A_{e-1}$ over $\mathbb{L}$, for
all $i=s+1, \ldots , n+m$.
\end{proof}

\bigskip

The next section is devoted to showing slightly more precise results
in the same spirit of the previous Theorem.

\section{Toward an equivalent explicit ODE system} \label{ODE}

In this section we make use of Theorem \ref{despeje general
dimension 0} in order to give more precise results on the number of
derivations required to obtain from a system of type (\ref{sistema
ampliado}) an equivalent explicit ODE system. In the first
subsection, we show that for $0$-dimensional systems there exists a
more accurate upper bound for the number of derivatives and the
order of the involved variables by means of an adequate modification
of the notion of  differentiation index. In the second subsection,
we consider the case of positive dimension for first-order systems.
Finally, we discuss some quantitative and symbolic algorithmic
aspects.

\subsection{Equivalent explicit ODE systems and an alternative notion of
differentiation index
for zero-dimensional systems} \label{sigma mono}

Throughout this subsection, our input system defined in
(\ref{sistema ampliado}) is considered to be $0$-dimensional and
further, it is  assumed that no variables $X$ appear (i.e. $r=m$ and
$n=0$). More explicitly, we consider the following  DAE system with
generic second members:
\begin{equation} \label{dimension0}
(\Sigma)\ :=\ \left\{
\begin{array}
[c]{ccl}%
g_1(U, \dot{U},\ldots ,U^{(e_1)}) &=& Y_1 \\
&\vdots &\\
g_r(U, \dot{U},\ldots,U^{(e_r)}) &=& Y_r \\
\end{array}
\right. ,
\end{equation}
where $U:=U_1,\ldots ,U_r$ and $Y:=Y_1,\ldots ,Y_r$ are sets of
differential indeterminates, $e_j\in \mathbb{N}_0$ for every $1\le j
\le r$, and $e:=\max \{e_1,\ldots,e_r\}$. We suppose $e\ge 1$.

For each $1\le i\le r$ we set $\epsilon_i$ for the maximum of the
orders of the variable $U_i$ in the polynomials $g_1,\dots, g_r$.
Since we assume a  differential $0$-dimensional situation,  all
variables must appear in the system; that is, we have that $0\le
\epsilon_i\le e$ for all $i$.

We introduce a new set of differential variables $Z:=Z_1,\ldots
,Z_r$ verifying:
\begin{equation} \label{nuevas variables}
U_i=Z_i^{(e-\epsilon_i)}
\end{equation}
for $i=1,\ldots,r$. Writing the input system $(\Sigma)$ in terms of
the indeterminates $Z$ we obtain a new DAE system:
\begin{equation} \label{dimension0modificado}
(\widetilde{\Sigma})\ :=\ \left\{
\begin{array}
[c]{ccl}%
g_1(Z, \dot{Z},\ldots ,Z^{(e)}) &=& Y_1 \\
&\vdots &\\
g_r(Z, \dot{Z},\ldots,Z^{(e)}) &=& Y_r \\
\end{array}
\right. .
\end{equation}
Note that each variable $Z_i$ appears at order exactly $e$ in at
least one of the equation in the system $(\widetilde{\Sigma})$, and
$e$ is the maximum of the orders of all variables appearing in
$(\widetilde{\Sigma})$.

We denote $\widetilde{\sigma}$ the differentiation index of the
system $(\widetilde{\Sigma})$. The number $\widetilde{\sigma}$ can
be regarded as an invariant of the DAE system $({\Sigma})$ and it
represents, in some sense, a more accurate version of the global
differentiation index introduced above, allowing us to distinguish
the variables more precisely according to their maximal orders (c.f.
the notion of \textit{structural index} introduced in
\cite{pantelides}, \cite{poulsen}).
We first show that the differentiation index $\sigma$ of
$({\Sigma})$ is always an upper bound for $\widetilde{\sigma}$:

\begin{lemma} \label{sigtilde}
The inequality $\widetilde{\sigma}\le \sigma$ holds.
\end{lemma}

\begin{proof} Let $\widetilde{G}$ be the
polynomials defining the system $(\widetilde{\Sigma})$, and let
$\widetilde{\Delta}$ be the differential ideal generated by the
differential polynomials $\widetilde{G}$ in the differential
polynomial ring $\mathbb{L}\{Z\}$.
 For each $p\in \mathbb{N}_0$, we set $\widetilde{A}_{p}$ for the polynomial ring
$\mathbb{L}[Z^{[p]}]$ and $\widetilde{\Delta}_p$ for the ideal
generated by the polynomials $\widetilde{G}^{[p-1]}$ in the ring
$\widetilde{A}_{p-1+e}$ (note that the maximum of the orders of the
polynomials $\widetilde G$ equals $e$).

Denote by ${\mathcal Z}$ the set of indeterminates $Z_i^{(\ell_i)}$
where $i=1,\ldots ,r$ and $0\le \ell_i< e-\epsilon_i$ (if
$e=\epsilon_i$ no derivatives of $Z_i$ appear in the set ${\mathcal
Z}$). The variables ${\mathcal Z}$ will be considered as algebraic
independent variables with respect to the original variables $U$ and
their successive derivatives. We also use the following standard
notation: if $A$ is a ring, $\mathfrak{A}\subset A$ an ideal and $X$
a set of indeterminates over $A$, $\mathfrak{A}[X]$ denotes the
ideal of $A[X]$ consisting of all the polynomials whose coefficients
belong to $\mathfrak{A}$.

In order to prove the inequality $\widetilde{\sigma}\le \sigma$,
applying Theorem \ref{sadik} to $(\widetilde \Sigma)$, it suffices
to show that the inclusion of ideals
\[
\widetilde{\Delta}\cap \widetilde{A}_{e-1} \subseteq
\widetilde{\Delta}_{\sigma}\cap \widetilde{A}_{e-1}
\]
holds.

Let $H\in \widetilde{\Delta}\cap \widetilde{A}_{e-1}$. Since $H\in
\widetilde{\Delta}$, there exists a sufficiently big index $q\in
\mathbb{N}$ such that $H\in \widetilde{\Delta}_q$. So, we can
suppose that $H$ is a polynomial combination involving only
variables $Z$ of order at most $q$ of polynomials of type
$\widetilde{G}^{(p)}_j$  with $p\le q$.

Replacing $Z_i^{(e-\epsilon_i+\ell)}$ with $U_i^{(\ell)}$, for
$i=1,\dots,r$ and each non negative integer $\ell$, we see that the
polynomial $H$ lies in the ideal $\Delta_q[{\mathcal Z}]\cap
\mathbb{L}[U_1^{[\epsilon_1-1]},\ldots
,U_r^{[\epsilon_r-1]}][{\mathcal Z}]$ (note that
$\mathbb{L}[U_1^{[\epsilon_1-1]},\ldots
,U_r^{[\epsilon_r-1]}][{\mathcal Z}]=\widetilde{A}_{e-1}$ and that
if $\epsilon_i=0$ neither $U_i$ nor its successive derivatives
appear in this polynomial ring). Since $\epsilon_i\le e$ for all
$i$, we conclude that $H\in \Delta_q[{\mathcal Z}]\cap
A_{e-1}[{\mathcal Z}]=(\Delta_q\cap A_{e-1})[{\mathcal Z}]$. Now,
due to identity (\ref{indexmin}) in the statement of Theorem
\ref{sadik}, we have that $\Delta_q\cap A_{e-1} \subseteq
\Delta_\sigma\cap A_{e-1}$. We infer that $H\in (\Delta_\sigma\cap
A_{e-1})[{\mathcal Z}]=\Delta_\sigma[\mathcal Z]\cap
A_{e-1}[{\mathcal Z}]$. The result follows observing that
$\Delta_\sigma[\mathcal Z]= \widetilde{\Delta}_\sigma$ and
$A_{e-1}[{\mathcal Z}]= \widetilde{A}_{e-1}$.
\end{proof}

\bigskip

The following pendulum-type system shows that the inequality in the
previous Lemma may be strict:

\vskip8pt

\noindent \textbf{Example\ } Consider the DAE system:
\[
({\Sigma})\ :=\ \left\{
\begin{array}
[c]{ccl}%
U_1^{(2)}+U_1U_3&=& Y_1 \\
U_2^{(2)}+U_2U_3&=& Y_2\\
U_1^2+U_2^2-1 &=& Y_3 \\
\end{array}
\right. .
\]
The corresponding matrix $\mathfrak{J}_{5,1}$ (see Definition
\ref{defimatricespi}) is

\[\left(
\begin{tabular}
[c]{ccccccccccccccc}%
$1$ & $0$ & $0$ & \multicolumn{1}{|c}{$0$} & $0$ & $0$ &
\multicolumn{1}{|c}{$0$} & $0$ & $0$ & \multicolumn{1}{|c}{$0$} &
$0$ & $0$ & \multicolumn{1}{|c}{$0$} & $0$ & $0$\\ $0$ & $1$ & $0$ &
\multicolumn{1}{|c}{$0$} & $0$ & $0$ & \multicolumn{1}{|c}{$0$} &
$0$ & $0$ & \multicolumn{1}{|c}{$0$} & $0$ & $0$ &
\multicolumn{1}{|c}{$0$} & $0$ & $0$\\ $0$ & $0$ & $0$ &
\multicolumn{1}{|c}{$0$} & $0$ & $0$ & \multicolumn{1}{|c}{$0$} &
$0$ & $0$ & \multicolumn{1}{|c}{$0$} & $0$ & $0$ &
\multicolumn{1}{|c}{$0$} & $0$ & $0$\\ \cline{1-3}\cline{3-3}
\\[-8mm]
&&&&&&\multicolumn{1}{|c}{} &&& \multicolumn{1}{|c}{} &&& \multicolumn{1}{|c}{} \\ %
$0$ & $0$ & $0$ & $1$ & $0$ & $0$ & \multicolumn{1}{|c}{$0$} & $0$ &
$0$ & \multicolumn{1}{|c}{$0$} & $0$ & $0$ &
\multicolumn{1}{|c}{$0$} & $0$ & $0$\\ $0$ & $0$ & $0$ & $0$ & $1$ &
$0$ & \multicolumn{1}{|c}{$0$} & $0$ & $0$ &
\multicolumn{1}{|c}{$0$} & $0$ & $0$ & \multicolumn{1}{|c}{$0$} & $0$ & $0$\\
$0$ & $0$ & $0$ & $0$ & $0$ & $0$ & \multicolumn{1}{|c}{$0$} & $0$ &
$0$ & \multicolumn{1}{|c}{$0$} & $0$ & $0$ &
\multicolumn{1}{|c}{$0$} & $0$ & $0$\\\cline{1-6} %
\\[-8mm]
&&&&&& &&& \multicolumn{1}{|c}{} &&& \multicolumn{1}{|c}{} \\ %
$U_{3}$ & $0$ & $U_{1}$ & $0$ & $0$ & $0$ & $1$ & $0$ & $0$ &
\multicolumn{1}{|c}{$0$} & $0$ & $0$ & \multicolumn{1}{|c}{$0$} & $0$ & $0$\\
$0$ & $U_{3}$ & $U_{2}$ & $0$ & $0$ & $0$ & $0$ & $1$ & $0$ &
\multicolumn{1}{|c}{$0$} & $0$ & $0$ & \multicolumn{1}{|c}{$0$} & $0$ & $0$\\
$2U_{1}$ & $2U_{2}$ & $0$ & $0$ & $0$ & $0$ & $0$ & $0$ & $0$ &
\multicolumn{1}{|c}{$0$} & $0$ & $0$ & \multicolumn{1}{|c}{$0$} &
$0$ &
$0$\\\cline{1-9}%
\\[-8mm]
&&&&&& &&&  &&& \multicolumn{1}{|c}{} \\ %
$3\dot{U}^{\ ^{\ }}_{3}$ & $0$ & $3\dot{U}_{1}$ & $U_{3}$ & $0$ &
$U_{1}$ & $0$ &
$0$ & $0$ & $1$ & $0$ & $0$ & \multicolumn{1}{|c}{$0$} & $0$ & $0$\\
$0$ & $3\dot{U}_{3}$ & $3\dot{U}_{2}$ & $0$ & $U_{3}$ & $U_{2}$ &
$0$ &
$0$ & $0$ & $0$ & $1$ & $0$ & \multicolumn{1}{|c}{$0$} & $0$ & $0$\\
$6\dot{U}_{1}$ & $6\dot{U}_{2}$ & $0$ & $2U_{1}$ & $2U_{2}$ & $0$ &
$0$ & $0$ & $0$ & $0$ & $0$ & $0$ & \multicolumn{1}{|c}{$0$} & $0$ &
$0$\\\cline{1-12}%
\\[-8mm]
&&&&&& &&&  &&& \\ %
$6U_{3}^{(2)}$ & $0$ & $6U_{1}^{(2)}$ & $4\dot{U}_{3}$ & $0$ &
$4\dot{U}_{1}$ & $U_{3}$ & $0$ & $U_{1}$ & $0$ & $0$ & $0$ & $1$ &
$0$ &
$0$\\
$0$ & $6U_{3}^{(2)}$ & $6U_{2}^{(2)}$ & $0$ & $4\dot{U}_{3}$ &
$4\dot{U}_{2}$ & $0$ & $U_{3}$ & $U_{2}$ & $0$ & $0$ & $0$ & $0$ &
$1$ &
$0$\\
$12U_{1}^{(2)}$ & $12U_{2}^{(2)}$ & $0$ & $8\dot{U}_{1}$ &
$8\dot{U}_{2}$ & $0$ & $2U_{1}$ & $2U_{2}$ & $0$ & $0$ & $0$ & $0$ & $0$ & $0$ & $0$%
\end{tabular}
\right).
\]

The dimension (over the field $k\langle U \rangle$) of the
corresponding kernels of $\mathfrak{J}^t_{i,1}$ are $\mu_i=i$, for
$i=0,\ldots ,4$, and $\mu_5=4$. Therefore, the differentiation index
of the system $(\Sigma)$ is $\sigma=4$.

Changing the variables as in (\ref{nuevas variables}): $Z_1:=U_1$,
$Z_2:=U_2$ and ${Z}_3^{(2)}=U_3$, we transform the previous system
into:
\[
(\widetilde{{\Sigma}})\ :=\ \left\{
\begin{array}
[c]{ccl}%
Z_1^{(2)}+Z_1Z_3^{(2)}&=& Y_1 \\
Z_2^{(2)}+Z_2Z_3^{(2)}&=& Y_2\\
Z_1^2+Z_2^2-1 &=& Y_3 \\
\end{array}
\right. .
\]
Here,  the corresponding Jacobian sub-matrix
$\widetilde{\mathfrak{J}}_{3,1}$ is:
\[\left(
\begin{tabular}
[c]{ccccccccc}%
$1$ & $0$ & $Z_{1}$ & \multicolumn{1}{|c}{$0$} & $0$ & $0$ &
\multicolumn{1}{|c}{$0$} & $0$ & $0$\\
$0$ & $1$ & $Z_{2}$ & \multicolumn{1}{|c}{$0$} & $0$ & $0$ &
\multicolumn{1}{|c}{$0$} & $0$ & $0$\\
$0$ & $0$ & $0$ & \multicolumn{1}{|c}{$0$} & $0$ & $0$ &
\multicolumn{1}{|c}{$0$} & $0$ & $0$\\\cline{1-3} & & & & & &
\multicolumn{1}{|c}{} \\[-3.5mm]  $0$ & $0$ & $\dot{Z}_{1}$ & $1$ & $0$ &
$Z_{1}$ & \multicolumn{1}{|c}{$0$}
& $0$ & $0$\\
$0$ & $0$ & $\dot{Z}_{2}$ & $0$ & $1$ & $Z_{2}$ &
\multicolumn{1}{|c}{$0$}
& $0$ & $0$\\
$0$ & $0$ & $0$ & $0$ & $0$ & $0$ & \multicolumn{1}{|c}{$0$} & $0$ &
$0$\\\cline{1-6} \\[-3.5mm]  $Z_{3}^{(2)}$ & $0$ & $Z_{1}^{(2)}$ & $0$ & $0$
& $2\dot{Z}_{1}$ & $1$ &
$0$ & $Z_{1}$\\
$0$ & $Z_{3}^{(2)}$ & $Z_{2}^{(2)}$ & $0$ & $0$ & $2\dot{Z}_{2}$ &
$0$ &
$1$ & $Z_{2}$\\
$2Z_{1}$ & $2Z_{2}$ & $0$ & $0$ & $0$ & $0$ & $0$ & $0$ & $0$%
\end{tabular}
\right),
\]
and the dimensions of the kernels of the matrices
$\widetilde{\mathfrak{J}}^t_{k,1}$ (for $k=0,1,2,3$) are $\mu_0=0$,
$\mu_1=1$, $\mu_2=2=\mu_3$. Therefore, the differentiation index of
the system $(\widetilde{{\Sigma}})$ is $\widetilde{\sigma}=2$, which
is smaller than the index $\sigma = 4$ of the original system
$(\Sigma)$.


\begin{remark} Applying a standard change of variables to the system $(\Sigma)$
defined in (\ref{dimension0}) we obtain an equivalent first-order system
$(\widehat{\Sigma})$ with differentiation index $\widehat{\sigma}$. With a
similar argument to that of Lemma \ref{sigtilde} it is easy to see that
$\widetilde{\sigma}\le \widehat{\sigma} \le \sigma$. In the previous example,
we have that $\widehat{\sigma}=3$,  which shows that the inequalities may be
strict.

\end{remark}

By means of the modified differentiation index $\widetilde{\sigma}$,
it is possible to give a more accurate version of Theorem
\ref{despeje general dimension 0}:

\begin{theorem} \label{despeje mejorado}
Let $(\Sigma)$ be a DAE system as in (\ref{dimension0}). For
$i=1,\ldots ,r$ we denote $\epsilon_i$ the maximum of the orders of
the variable $U_i$ in the equations defining $(\Sigma)$. Then there
exists a set $\mathcal{U}\subset \{ U_1^{[\epsilon_1-1]},\ldots
,U_r^{[\epsilon_r-1]}\}$ (here, if $\epsilon_i=0$, no variable
$U_i^{(p)}$ appears) and separable polynomials $P_1,\ldots ,P_r$
such that
\[P_i(Y^{[\widetilde{\sigma}]},\mathcal{U},U_i^{(\epsilon_i)})\in
({G}_1^{[\widetilde{\sigma}]}, \ldots,
{G}_r^{[\widetilde{\sigma}]}).\] In other words, there exist
non-trivial (separable) polynomial relations between each
$U_i^{(\epsilon_i)}$ and a fixed family of derivatives
$\mathcal{U}\subset \{ U_1^{[\epsilon_1-1]},\ldots
,U_r^{[\epsilon_r-1]}\}$ (which is a family of order strictly lower
than $\max_i\{\epsilon_i\}-1$) that can be obtained from the first
$\widetilde{\sigma}$ many derivatives of the input equations.
\end{theorem}

\begin{proof}
Let $(\widetilde{\Sigma})$ be the DAE system obtained from
$(\Sigma)$ after the change of variables (\ref{nuevas variables}).
For $j=1,\dots, r$, denote by $\widetilde{G_j}$ the polynomial
obtained from $G_j$ after this change.

Applying Theorem \ref{despeje general dimension 0} to
$(\widetilde{\Sigma})$, there exists a subset $\Xi \subset \{
Z_1^{[e-1]},\ldots ,Z_r^{[e-1]}\}$ such that for $i=1,\ldots ,r$
there exists a separable polynomial $Q_i$ satisfying
\begin{equation} \label{implicita1}
Q_i(Y^{[\widetilde{\sigma}]},\Xi ,Z_i^{(e)})\in
(\widetilde{G}^{[\widetilde{\sigma}]})\subset
k[Y^{[\widetilde{\sigma}]},Z^{[\widetilde{\sigma}+e]}].
\end{equation}

Fix an index $i$ and its polynomial $Q_i$. Writing $U_j^{(\ell)}$
for each variable $Z_j^{(e-\epsilon_j+\ell)}$, we have that the set
$\Xi$ may be decomposed as $\Xi=\Xi_U\,\cup \, \Xi_Z$ where $\Xi_U
\subset \{ U_1^{[\epsilon_1-1]},\ldots ,U_r^{[\epsilon_r-1]}\}$ and
$\Xi_Z\subset \{ Z_1^{[e-\epsilon_1-1]},\ldots ,
Z_r^{[e-\epsilon_r-1]} \}$. With this notation, condition
$(\ref{implicita1})$ can be written as
$Q_i(Y^{[\widetilde{\sigma}]},\Xi_U,\Xi_Z ,U_i^{(\epsilon_i)})\in
({G}^{[\widetilde{\sigma}]}), $ where the ideal
$({G}^{[\widetilde{\sigma}]})$ is considered in the polynomial ring
$k[Y^{[\widetilde{\sigma}]},Z_1^{[e-\epsilon_1-1]},\ldots,
Z_r^{[e-\epsilon_r-1]}, U_1^{[\widetilde{\sigma}+\epsilon_1]},\ldots
,U_r^{[\widetilde{\sigma}+\epsilon_r]}]$.

Then, the Theorem follows from (\ref{implicita1}) taking $\mathcal
U:=\Xi_U$ and evaluating the variables $Z_j^{[e-\epsilon_j-1]}$ for
$j=1,\dots, r$ in generic elements of $k$.
\end{proof}

\bigskip

We remark that Theorem \ref{despeje mejorado} improves upon Theorem
\ref{despeje general dimension 0} in at least two points: first, the
number of derivatives required to obtain the polynomial relations is
the modified differentiation index $\widetilde{\sigma}$ instead of
$\sigma$. On the other hand, the order of each variable $U_i$ in the
corresponding relation $P_i$ is the maximum order of the
\textit{single} variable $U_i$ in the system (namely, $\epsilon_i$)
instead of the maximum of the orders of \textit{all} variables
(namely, $e$).

\subsection{Differential
transcendence bases and equivalent ODE form for generic first-order
DAE systems of positive dimension} \label{implicit}

In the previous sections (more precisely, in Theorems \ref{despeje
general dimension 0} and \ref{despeje mejorado}) we saw how to
obtain implicit equations for derivatives of low order of each
variable in a DAE system after at most $\sigma$ (or
$\widetilde{\sigma}$) many differentiations of the input equations.
In both results a ``square" $0$-dimensional situation was assumed.

Here we show a result of the same kind in the case of
\textit{positive} differential dimension for generic
\textit{first-order} systems of type (\ref{sistema ampliado}). Even
though the straightforward idea  could be applied (namely, the
localization into a differential transcendence basis to reduce the
problem to a $0$-dimensional situation), different transcendence
bases may lead to systems with different quantitative properties.
For instance, the number of derivatives required to obtain an
equivalent explicit ODE system may change. In this sense, there
exist transcendence bases which are more adequate for our global
analysis. Before stating the precise results we consider a simple
example in order to illustrate this kind of phenomena.

\subsubsection{Example} \label{example}

Consider the following $1$-dimensional first-order DAE system with
coefficients in $k:=\mathbb{Q}$ borrowed essentially from
\cite[Section 3.4]{Fliess}:
\begin{equation} \label{ejemplo 1} \left\{
\begin{array}
[c]{ccl}%
Y_1 & = & U_1+\dot{U}_m\\
Y_2 & = & U_2+\dot{U}_1\\
 &\vdots  & \\
Y_{m-1} & = & U_{m-1}+\dot{U}_{m-2}
\end{array}
\right.
\end{equation}
with $m>2$.

Here, we have $n=0$ and $r=m-1$. The associated
field $\mathbb{K}$ is $\mathbb{Q}\langle U\rangle$ and the matrix
$\mathfrak{J}_{1,0}\in \mathbb{K}^{(m-1)\times m}$ is
\[ \left(
\begin{array}{ccccc}
0 & \cdots  & \cdots  & 0 & 1 \\
1 & 0 & \cdots  & 0 & 0 \\
\vdots  & \ddots  & \ddots  & \vdots  & \vdots  \\
0 & \cdots  & 1 & 0 & 0%
\end{array}%
\right),
\]
which has full row rank $m-1$. Hence, we have $\mu_{1}=\mu_{0}=0$
and so, Definition \ref{indice} states that the differentiation
index $\sigma$ of the system (\ref{ejemplo 1}) is equal to $0$.

Following the slightly informal definition of the differentiation
index given in the Introduction, the fact that $\sigma=0$ should
imply that all the derivatives $\dot{U}_{i}$ can be written in terms
of the variables $U_1,\ldots ,U_m$ using the equations (i.e., the
variables $Y_1,\dots, Y_{m-1}$). While this is trivially true for
all the variables different from $U_{m-1}$,
\begin{equation} \label{despeje1}
\begin{array}
[c]{ccl}%
\dot{U}_m & = & Y_1-{U}_1\\
\dot{U}_1 & = & Y_2-{U}_2\\
\ & \vdots & \ \\
\dot{U}_{m-2} & = & Y_{m-1}-{U}_{m-1},
\end{array}
\end{equation}
it is not so clear how to find a similar relation for
$\dot{U}_{m-1}$, because it does not appear in the equations.

It is quite natural to consider $U_{m-1}$
as a free variable and then to interpret the previous system as a $0$-dimensional
system over the field $k_1:=\mathbb{Q}\langle U_{m-1} \rangle$. Now, the matrix
$\mathfrak{J}_{1,0}$ is the $(m-1)$-square matrix
\[
\left(
\begin{array}
[c]{cccc}%
0 & \cdots & 0 & 1\\
1 & \cdots & 0 & 0\\
\vdots & \ddots & \vdots & \vdots\\
0 & \cdots & 1 & 0
\end{array}
\right),
\]
which is clearly non singular. Hence $\mu_{1}=\mu_{0}=0$ and so, the
new associated differentiation index $\sigma_1$ is equal to $0$. The
relations (\ref{despeje1}) can be seen now as a full rewriting of
the derivatives in terms of the variables (c.f. Theorem \ref{despeje
general dimension 0}). Furthermore, we observe that the order of the
ideal associated to the system does not change by the extension of
the ground field from $\mathbb{Q}$ to $\mathbb{Q}\langle U_{m-1}
\rangle$: following Remark \ref{indice-hilbert}, we have
$\textrm{ord}_{\mathbb{L}\langle U_{m-1}\rangle}(\Delta)
=\textrm{ord}_{\mathbb{L}}(\Delta)=m-1$, since the sequence
$(\mu_{k})_k$ is the same in both cases.


On the other hand, if in the system (\ref{ejemplo 1}) we take $U_m$
as the free variable and consider the original system as a
$(m-1)$-square $0$-dimensional DAE system over the ground
differential field $k_2:=\mathbb{Q}\langle U_m\rangle$, the matrix
$\mathfrak{J}_{1,0}$ is the $(m-1)$-square matrix
\[ \left(
\begin{array}{cccc}
0 & \cdots  & \cdots  & 0  \\
1 & 0 & \cdots  & 0  \\
\vdots  & \ddots  & \ddots  & \vdots  \\
0 & \cdots  & 1 & 0
\end{array}%
\right).
\]
So, $\mu_{1}\ne0=\mu_{0}$ and therefore, the differentiation index
$\sigma_2$ of the system (\ref{ejemplo 1}) over $k_2$ is strictly
positive. In fact, it is easy to see that $\sigma_2=m-1$.  Now,
Theorem \ref{despeje general dimension 0} ensures that the order of
derivatives of the equations required to express the derivatives
$\dot{U}_1,\ldots,\dot{U}_{m-1}$ in terms of $U_1,\ldots,U_{m-1}$ is
at most $m-1$. Indeed, the variables $U_1,U_2,\ldots ,U_{m-1}$ can
be written in terms of derivatives of the equations (i.e.
derivatives of the variables $Y_1,\dots, Y_{m-1}$) and elements of
the base field $k_2$ as follows:
\begin{eqnarray*}
U_1 & = & Y_1-\dot{U}_m\\
U_2 & = & Y_2-\dot{Y}_1+{U}^{(2)}_m\\
U_3 & = & Y_3-\dot{Y}_2+{Y}^{(2)}_1-{U}^{(3)}_m\\
\ & \vdots & \ \\
U_{m-1} & = & Y_{m-1}-\dot{Y}_{m-2}+\cdots +(-1)^{m-2}
Y^{(m-2)}_1+ (-1)^{m-1}U^{(m-1)}_m.
\end{eqnarray*}
Observe also that the order of the ideal is not preserved after
localization in $U_m$, since $\textrm{ord}_{\mathbb{L}\langle
U_m\rangle}(\Delta)=0 \ne m-1 = \textrm{ord}_{\mathbb{L}}(\Delta)$.

\bigskip

The previous example shows two different situations arising when
considering different transcendence bases. The first localization of
the system (\ref{ejemplo 1}) seems to follow the behavior of the
original system more closely than the second one, since in that case
the derivatives of the unknowns which are not in the transcendence
bases could be written in terms of the unknowns themselves using as
many derivatives of the equations as the differentiation index of
the original system.

In the next two subsections we will show that for any first-order
system of type (\ref{sistema ampliado}) there are suitable
differential transcendence bases which enable us to obtain relations
between the remaining differential dependent variables using ``few''
(as many as the differentiation index) derivatives of the equations.

\subsubsection{Differential transcendence basis preserving the order}

In the sequel, we suppose that the input system $(\Sigma)$ is of
first order (or equivalently, $e=1$). The notations correspond to
those introduced in the previous sections.

We will denote $\cF_i:= {\rm Frac}(A_i/ (\Delta \cap A_i))$ for
every $i\in \N_0$. As before, we will use the same notation for an
element of $A_i$ or its class in $\cF_i$ whenever the ring in which
it is considered is clear from the context. The fact that $\cF_i
\hookrightarrow \cF_{i+1}$ for every $i\in \N_0$ allows us to
consider any subset of $\cF_i$ as a subset of $\cF_{j}$ for every
$j\ge i$, which will also be done without changing notations.

\begin{lemma}\label{algebraicos}
Let $\cB \subset A_i$ and let $\zeta \in A_i$ be a polynomial such
that its class $\zeta \in \cF_i$ is algebraic over
$\L(\cB)$. Then, $\dot \zeta \in \cF_{i+1}$
is algebraic over $\L(\cB \cup \dot \cB)$,
where $\dot \cB$ denotes the set of classes of all derivatives of
elements in $\cB$. In particular, if $\L(\cB) \hookrightarrow
\cF_i$ is an algebraic field extension, then $\L(\cB \cup \dot
\cB) \hookrightarrow \cF_{i+1}$ is also algebraic.

\end{lemma}

\begin{proof} The result is immediate if $\zeta \in \cB$. So, let
us consider the case when $\zeta \notin \cB$. Let $P \in
\L(\cB)[T]$ be the minimal polynomial of $\zeta$ with respect to
the field extension
$\L(\cB)\hookrightarrow \cF_i$. Multiplying it by a non-zero
element in $\L[\cB]$, we may assume that $P\in \L[\cB,T]$ and has
non-zero leading coefficient.

We have $P(\cB, \zeta) \in \Delta \cap A_i$, and so $\dot P(\cB
\cup \dot \cB, \zeta, \dot \zeta) \in \Delta \cap A_{i+1}$. Now,
$\dot P(\cB \cup \dot \cB, \zeta, \dot \zeta) = Q(\cB \cup \dot
\cB, \zeta) + \frac{\partial P}{\partial T}(\cB, \zeta) \, \dot
\zeta$ for some polynomial $Q$. As $\deg_T(\frac{\partial
P}{\partial T}) < \deg_T(P)$, the minimality of $P$ implies that
$\frac{\partial P}{\partial T}(\cB, \zeta)\notin \Delta$ and so,
$\dot P(\cB \cup \dot \cB, \zeta, T)$ is a non-zero polynomial in
$\L(\cB \cup\dot \cB, \zeta)[T]$ annihilating $\dot \zeta$ in
$\cF_{i+1}$. This implies that $\dot \zeta$ is algebraic over
$\L(\cB \cup \dot \cB, \zeta)$.

Since the field sub-extension $\L(\cB \cup \dot \cB)
\hookrightarrow \L(\cB \cup \dot \cB, \zeta)$ of $\L(\cB \cup
\dot \cB) \hookrightarrow \cF_{i+1}$ is algebraic, we conclude
that $\dot \zeta$ is algebraic over $\L(\cB \cup \dot \cB)$.
\end{proof}

\begin{proposition} \label{buenabase1} Let $s:= {\rm
ord}_\L(\Delta)$. There exists disjoint subsets $W:=\{ W_1, \dots, W_{m-r}\}$
and $ \xi:=\{\xi_1,\dots, \xi_s\}$ of the set $\{X_1,\dots, X_n, U_1,\dots,
U_m\}$ such that $\cB_i:=\{W_1^{[i]}, \dots , W_{m-r}^{[i]},\xi_1,\dots,
\xi_s\}$ is a transcendence basis of the algebraic field extension $\L
\hookrightarrow \cF_i$ for every $i\in \N_0$. In particular, the set $\{
W_1,\dots, W_{m-r}\}$ is a differential transcendence basis of the differential
field extension $\L \hookrightarrow {\rm Frac}(\L\{X, U\} /\Delta)$.
\end{proposition}

\begin{proof} Let $\cB_0\subset \{X_1,\dots, X_n, U_1, \dots, U_m\}$ be a
transcendence basis of $\L \hookrightarrow \cF_0$. Then,
$\L(\cB_0) \hookrightarrow \cF_0$ is an algebraic field extension
and so, due to Lemma \ref{algebraicos}, $\L(\cB_0 \cup \dot
\cB_0) \hookrightarrow \cF_1$ is also algebraic. Hence, $\cB_0
\cup \dot \cB_0$ contains a transcendence basis of $\cF_1$ over
$\L$. Since $\cB_0$ is algebraically independent over $\L$, and
${\rm trdeg}_\L(\cF_1) = m-r + {\rm trdeg}_\L(\cF_0)$ (see the proof of
Theorem \ref{hilbert constante}), there exists a subset
$\widetilde\cB_0 \subset \dot \cB_0 $ with $m-r$ elements such
that $\cB_1:= \cB_0 \cup \widetilde \cB_0$ is a transcendence
basis of the extension $\L \hookrightarrow \cF_1$.

Let us denote $W_1,\dots, W_{m-r}$ the variables whose first
derivatives are all the elements in $\widetilde \cB_0$ (note that
$\{W_1,\dots, W_{m-r}\} \subset \cB_0$) and let $\{\xi_1,\dots,
\xi_s\}:= \cB_0 \setminus \{W_1,\dots, W_{m-r}\}$ (observe that
$\# \mathcal{B}_0=(m-r)+s$ since the Hilbert-Kolchin function
coincide with its associated polynomial as shown in Theorem
\ref{hilbert constante}). We will show that, for every $i\in \N$,
the set $\cB_i:=\{W_1^{[i]}, \dots, W_{m-r}^{[i]}, \xi_1, \dots,
\xi_s\}$ is a transcendence basis of $\L \hookrightarrow \cF_i$.

The case when $i=1$ follows from our previous construction. Let
us assume now that  $\cB_i$ is a transcendence basis of $\L
\hookrightarrow \cF_i$ for a fixed positive integer $i\in \N$.
Then, by Lemma  \ref{algebraicos}, $\L(\cB_i\cup \dot \cB_i) =
\L(\cB_{i+1} \cup \{\dot \xi_1,\dots, \dot \xi_s\})
\hookrightarrow \cF_{i+1}$ is an algebraic field extension. Now,
$\L(\cB_{i+1}) \hookrightarrow \L(\cB_{i+1} \cup \{\dot
\xi_1,\dots, \dot \xi_s\})$ is an algebraic sub-extension of
$\L(\cB_{i+1})\hookrightarrow \cF_{i+1}$, since each of the
elements $\dot \xi_j$ is obviously algebraic as an element of
$\L(\cB_{i+1})$ over $\L(\cB_1)$. Therefore, $\L(\cB_{i+1})
\hookrightarrow \cF_{i+1}$ is an algebraic extension and, taking
into account that ${\rm trdeg}_\L(\cF_{i+1})$ equals the
cardinality of $\cB_{i+1}$ (see Theorem \ref{hilbert constante}),
we conclude that $\cB_{i+1}$ is a transcendence basis of $\L
\hookrightarrow \cF_{i+1}$.
\end{proof}

\bigskip

It is clear from the proof that the differential transcendence basis
$\{W_1,\ldots,W_{m-r}\}$  preserves the order of the ideal (the
constant term of the Hilbert-Kolchin polynomial) after extending the
ground field $\mathbb L$ to $\mathbb{L}\langle W_1,\ldots
,W_{m-r}\rangle$.

\subsubsection{An equivalent ODE system}\label{psyqs}

Theorem \ref{hilbert constante} and Proposition \ref{buenabase1}
enable us to deduce the following ``implicit function type" result
in terms of the differentiation index $\sigma$ introduced in
Definition \ref{indice} (see also \cite[Section 3]{Fliess}), an
analogue of Theorem \ref{despeje general dimension 0} for
first-order non-square generic DAE systems:

\begin{corollary} \label{implicita}
The variables $X,U$ can be split into three subsets $W:=\{W_1,\dots,
W_{m-r}\}$, $\xi:=\{\xi_1,\ldots ,\xi_s\}$ and $\eta:=\{\eta_{s+1},\ldots
,\eta_{n+r}\}$ so that:
\begin{enumerate}
\item $W$ is a differential transcendence basis of $\L \hookrightarrow {\rm
Frac}(\L\{X, U\} / \Delta)$;
 \item $W^{[j]} \cup \xi$ is an algebraic
transcendence basis of $\L\hookrightarrow \cF_j$ for all $j\in \mathbb{N}_0$;
\item for each $i=1,\ldots ,s$ there exists a non-zero separable polynomial
$P_i$ with coefficients in the base field $k$, such that $P_i(Y^{[\sigma]},W,
\dot{W}, \xi,  \dot{\xi}_i)\in (F^{[\sigma]}, G^{[\sigma]})\subset
k[Y^{[\sigma]}, X^{[\sigma+1]}, U^{[\sigma + 1]}]$;
\item 
for each $i=s+1,\ldots ,n+r$, there exists a non-zero separable polynomial
$P_i$ with coefficients in the base field $k$, such that $P_i(Y^{[\sigma-1]},W,
\xi,  \eta_i)\in (F^{[\sigma-1]}, G^{[\sigma-1]})\subset k[Y^{[\sigma-1]},
X^{[\sigma]}, U^{[\sigma]}]$.
\end{enumerate}
\end{corollary}

\begin{proof} Let $W$ and $\xi$ be subsets of variables as in
Proposition \ref{buenabase1}. Then, the first and second conditions in the
statement hold. Let $\eta:=\{\eta_{s+1},\ldots ,\eta_{n+r}\} $ be the set of
the remaining unknowns $X,U$ (i.e., those different from the $W$'s and the
$\xi$'s).

Now the proof runs \textit{mutatis mutandis} as in Theorem \ref{despeje general
dimension 0}. For instance, for the fourth item we observe that  for every $i$,
$s+1\le i \le n+r$, the element $\eta_i \in \cF_0$ is algebraic over $\L(W \cup
\xi)$, since $W \cup \xi$ is a transcendence basis of $\L \hookrightarrow
\cF_0$. Hence, there is a non-zero polynomial $\widehat{P}_i$ with coefficients
in $\mathbb{L}$ such that $\widehat{P_i}(W, \xi, \eta_i)\in \Delta\cap A_0 =
\Delta_\sigma \cap A_0$ (see Theorem \ref{sadik}).  In the same way, for the
third item we have that  $W\cup \dot W \cup \xi$ is a transcendence basis of
$\L \hookrightarrow \cF_1$ which ensures the existence of non-zero polynomials
$\widehat{P}_i$ $(i=1,\dots, s)$ with coefficients in $\mathbb{L}$ verifying
$\widehat{P}_i(W, \dot W, \xi, \dot{\xi}_i)\in \Delta \cap  A_1 =
\Delta_{\sigma+1}\cap A_1$.

The polynomials $P_i$'s  that we are looking for can be easily obtained from
the previous $\widehat{P}_i$'s, by multiplying them by adequate factors in $k\{
Y \}$ and evaluating the superfluous variables $Y^{(l)}$ at suitably chosen
elements of the base field $k$ (for all $l\ge \sigma$ in the case of $i\ge s+1$
 and for all $l\ge \sigma+1$ for $i\le s$).
\end{proof}

\bigskip

Formally, the fourth statement of the previous Corollary makes no sense if the
differentiation index $\sigma$ is zero. However, it admits a natural
interpretation also in this case: this situation corresponds exactly to the
case where the Theorem of Implicit Functions can be applied in order to write
each derivative of the variables $\{X,U\}\setminus\{W\}$ in terms of the same
variables and derivatives of the $W$'s (obviously not in a polynomial nor a
rational way). So, the fourth item must be empty. More precisely:

\begin{remark} Under the conditions of Corollary \ref{implicita}, suppose
also that the differentiation index $\sigma$ is zero. Following Remark
\ref{indice-hilbert}, the Hilbert-Kolchin polynomial of $\Delta$ over
$\mathbb{L}$ is $\mathcal{H}_{\Delta}(T)=(m-r)(T+1)+(n+r)$ and the order of
this ideal is $n+r$ (recall that we assume $e=1$ and $\mu_0$ is defined to be
$0$). So,  the set $\eta$ is empty, or equivalently $ \xi=\{X,U\}\setminus
\{W\}$. Then, there exist non-zero polynomials $P_i$, $i=1,\ldots, n+r$, such
that $P_i(Y,W, \dot W, \xi,\dot{\xi}_i)\in (F, G)\subset k[Y, X, \dot{X}, U,
\dot{U}]$.
\end{remark}

\subsection{Quantitative and algorithmic aspects}
\label{quantitative}

\subsubsection{Degree bounds of the implicit equations} \label{grados}

The non-zero polynomials $P_i$ of Theorems \ref{despeje general dimension 0} \&
\ref{despeje mejorado} and Corollary \ref{implicita}, are not uniquely
determined without additional requirements (as minimality of order and degree,
irreducibility, etc.). However, the conditions stated in those results allow us
to choose a family of such polynomials that can be regarded as eliminating
polynomials of suitable algebraic-geometric situations, which enables us to
estimate their degrees.

In order to illustrate these facts consider for instance the
situation of Corollary \ref{implicita}: let $\bar {k}$ be a fixed
algebraic closure of the ground field $k$. For each $N\in
\mathbb{N}$ we denote by $\mathbb{A}^N$ the affine space $\bar{k}^N$
equipped with the Zariski topology. Set $N_0:=(r+n+m)\sigma+ n+m$
and $N_1:=(r+n+m)(\sigma +1)+n+m$ and let $\mathbb{V}_0\subset
\mathbb{A}^{N_0}$ and $\mathbb{V}_1\subset \mathbb{A}^{N_1}$ be the
algebraic varieties defined by the ideals $(F^{[\sigma-1]},
G^{[\sigma-1]})\subset k[Y^{[\sigma-1]}, X^{[\sigma]},
U^{[\sigma]}]$ and $(F^{[\sigma]}, G^{[\sigma]})\subset
k[Y^{[\sigma]}, X^{[\sigma+1]}, U^{[\sigma+1]}]$ respectively, that
is: \begin{equation} \label{variedades}
\mathbb{V}_0:=\{
F^{[\sigma-1]}=0\ ,\ G^{[\sigma-1]}=0\}\qquad \textrm{and}\qquad
\mathbb{V}_1:=\{ F^{[\sigma]}=0\ ,\ G^{[\sigma]}=0 \}.\end{equation}
Note that both varieties are irreducible complete intersection and
their dimensions are $m(\sigma +1)+n$ and $m(\sigma +2)+n$
respectively.

Let  $W:= \{ W_1,\dots, W_{m-r}\}$, $\xi:=\{ \xi_1,\ldots ,\xi_s \}$ and
$\eta:=\{\eta_{s+1},\ldots ,\eta_{n+r}\}$ be a partition of the set of
variables $\{X, U\}$ as in Corollary \ref{implicita}. We define linear
projections $\theta_i$ ($i=1, \ldots, s$) and $\pi_i$ ($i=s+1, \ldots, n+r$) as
follows:
\[\begin{array}{rlclc}
\theta_i:&\mathbb{V}_1\rightarrow \mathbb{A}^{r(\sigma -1)+2m+s+1}, & &
\theta_i(y^{[\sigma]},x^{[\sigma+1]},u^{[\sigma+1]}):=(y^{[\sigma]},w, \dot w,
\xi,\dot{\xi}_i);\\[2mm]
\pi_i:&\mathbb{V}_0\rightarrow \mathbb{A}^{r\sigma+m-r+s+1}, & &
\pi_i(y^{[\sigma-1]},x^{[\sigma]},u^{[\sigma]}):=(y^{[\sigma-1]},w,\xi,\eta_i).
\end{array}
\]

{}From Proposition \ref{buenabase1}, we deduce that the set $\{Y^{[\sigma]},W,
\dot W, \xi\}$  (resp.~$\{Y^{[\sigma-1]},W,\xi\}$) is algebraically independent
in the fraction field $k(\mathbb{V}_1)$ (resp.~$k(\mathbb{V}_0)$) over $k$. On
the other hand, due to Corollary \ref{implicita}, the set $\{Y^{[\sigma]},W,
\dot W,\xi, \dot{\xi_i} \}$  (resp.~$\{Y^{[\sigma-1]},W, \xi,\eta_i\}$) is
algebraically dependent. Thus, the closure of the image of the map $\theta_i$
(resp.~$\pi_i$) is a $k$-definable irreducible hypersurface in the
corresponding ambient space and so, it can be defined by a single polynomial
lying in the ideal $(F^{[\sigma]}, G^{[\sigma]})$ (resp. $(F^{[\sigma-1]},
G^{[\sigma-1]})$) whose total degree (see \cite[Lemma 2]{Heintz}) is bounded by
$\deg\mathbb{V}_1$ (resp.~$\deg\mathbb{V}_0$).

Applying the Bezout Inequality (see \cite[Theorem 1]{Heintz}) to obtain an
upper bound for $\deg\mathbb{V}_1$ (resp.~$\deg\mathbb{V}_0$), we conclude that
there exist polynomials $P_i$, for $i=1, \ldots , s$ (resp $i=s+1, \ldots ,
n+r$) meeting the conditions of Corollary \ref{implicita} whose total degrees
can be bounded by $\deg{P_i}\le d^{(\sigma+1)(n+r)}$ (resp.~$\deg{P_i}\le
d^{\sigma(n+r)}$), where $d$ denotes an upper bound for the total degree of the
polynomials in the input system.

\bigskip

A similar result can be obtained in the case of Theorem \ref{despeje
general dimension 0}  by considering the linear projections
\[ (y^{[\sigma]},x^{[\sigma+1]},u^{[\sigma +e]})\mapsto (y^{[\sigma]},\Xi,\xi_i^{(e)}), \]
\[(y^{[\sigma-1]},x^{[\sigma]},u^{[\sigma +e-1]})\mapsto (y^{[\sigma-1]},\Xi,\eta_i^{(e-1)}),\]
whose domains are the irreducible varieties $\mathbb{V}_1$ and
$\mathbb{V}_0$ defined in (\ref{variedades}), respectively.

Analogously, the projections which allow us to estimate the degrees
of the polynomials in Theorem \ref{despeje mejorado} are
\[ (y^{[\widetilde{\sigma}]},u^{[\widetilde{\sigma}+e]})\mapsto
(y^{[\widetilde{\sigma}]},\mathcal{U},u_i^{(\epsilon_i)}),\]
for $i=1,\ldots ,r$, all of them defined over the irreducible
variety
\begin{equation} \label{variedades2}
\widetilde{\mathbb{V}}_1:=\{G^{[\widetilde{\sigma}]}=0\}.
\end{equation}

%

\subsubsection{Algorithmic Issues} \label{issues}

This section presents algorithmic procedures for the computation of
the following objects: \begin{enumerate} \item[(1)]  The
differentiation index of the system (\ref{sistema ampliado}).
\item[(2)] A differential transcendence basis of a first-order
system of type (\ref{sistema ampliado}) preserving the order.
\item[(3)] The implicit relations in separated variables given by
Theorems \ref{despeje general dimension 0} \& \ref{despeje mejorado}
and Corollary \ref{implicita}.
\end{enumerate}

For algorithmic reasons, we will assume that our base field $k$ is
either $\Q$ or $\Q(t)$ with the usual derivation (in the case when
$k = \Q(t)$, the input polynomials will be assumed to have
coefficients in $\Q[t]$).

We start with a brief description of the computational model.

\subsubsection*{Basic algorithmic notions}

The objects our algorithms deal with are multivariate polynomials
which will be encoded by means of \emph{straight-line programs}
(i.e., arithmetic circuits which enable us to evaluate them at
any given point). The number of instructions in the program is
called the \emph{length} of the straight-line program. For a
brief description of the algorithmic model and the data structure
we will use, we refer the reader to \cite[Section 2.2]{djs} and
the references therein.

The basic subroutine we use is a polynomial-time probabilistic procedure for
the computation of the rank of a matrix with polynomial entries.  Roughly
speaking, the problem is reduced to the computation of the rank of a matrix
with entries in $\Q$ by randomly choosing integer values for the variables and
evaluating all matrix entries at them (see \cite[Lemma 24]{djs} for the error
probability analysis).

Our algorithms take as input a straight-line program of length $L$
encoding the polynomials $f_1,\dots, f_n$, $g_1, \dots, g_r$
appearing in the system (\ref{sistema ampliado}). However, the
intermediate computations involve not only these polynomials but
also their successive derivatives and so, straight-line programs
for these derivatives are needed as well. The existence of
``short'' straight-line programs encoding them is ensured by
\cite[Lemma 21]{djs} (see also \cite[Section 5.2]{MS}).

We recall that all our computations should be performed over the differential
field $\K$ with the derivation induced by $\dot X_i = f_i$ $(i=1,\dots, r)$
but, since the number of derivatives involved in each computation is
controlled, they can be achieved over the polynomial rings $k[X, U^{[l]}]$ for
adequate choices of $l$. We assume that an upper bound $d\in \N$ for the
degrees of the polynomials $f_1,\dots, f_n$, $g_1, \dots, g_r$ is known. In
order to estimate complexities and error probabilities, we also need upper
bounds for the degrees of the polynomials obtained by successive
differentiation of the input polynomials (i.e., those giving the isomorphism
between ${\rm Frac}(k\{Y, X, U\}/\Delta)$ and $\K$), which can be found in
\cite[Notation 6 and Remark 25]{djs}.

\subsubsection*{Computation of the differentiation index}

According to Definition \ref{indice} and Notation \ref{sacar i}, we
have $\sigma:= \min \{ k\in \N_0 \,/\, \mu_{k} = \mu_{k+1}\}$. This
minimum is obtained by computing and comparing the ranks of the
matrices $\fJ_{k,e-1}$ (which are computed over polynomial rings)
for successive values of $k\in \N$. The algorithm finishes, since we
have an a priori upper bound for $\sigma$ (see (\ref{remark
indice})).

The previous computation of $\sigma$ can be achieved with error
probability bounded by $\varepsilon$ within a complexity which is
polynomial in $n, m, r$, and linear in $\log d$, $\log \varepsilon$
and $L$ (recall that $d$ and $L$ are upper bounds for the degrees
and the size of the straight-line program representation of the
input polynomials). Note that this algorithm can also be applied
 for the computation of the modified index
$\widetilde{\sigma}$ introduced in Subsection \ref{sigma mono} with
the same complexity bounds.

\subsubsection*{Computation of a differential transcendence basis for
first-order systems}

The algorithmic computation of a differential transcendence basis of
the differential field extension $\L \hookrightarrow {\rm
Frac}(\L\{X, U\} /\Delta)$ preserving the order after localization
follows the procedure underlying the proof of Proposition
\ref{buenabase1}:
\begin{itemize}
\item compute a
transcendence basis $\cB_0\subset \{X_1,\dots, X_n, U_1, \dots,
U_m\}$ of $\L \hookrightarrow \cF_0$;
\item choose a subset
$\widetilde\cB_0 \subset \dot \cB_0 $ with $m-r$ elements such
that $\cB_1:= \cB_0 \cup \widetilde \cB_0$ is a transcendence
basis of $\L \hookrightarrow \cF_1$.
\end{itemize}
Then, the variables $W_1, \dots, W_{m-r}$ whose derivatives lie in
$\widetilde \cB_0$ form a differential transcendence basis of $\L
\hookrightarrow {\rm Frac}(\L\{X, U\} /\Delta)$ with the required
property.

The set $\cB_0$ is constructed recursively by adding one variable at a time. In
order to determine whether a subset of variables in $\cF_0$ (resp.~$\cF_1$) is
transcendental over the field $\L$, we use the fact that $\cF_0 \hookrightarrow
A_\sigma / \Delta _\sigma$ (resp.~$\cF_1 \hookrightarrow A_{\sigma +1}/
\Delta_{\sigma+1}$). Thus, the problem amounts to determine whether a subset of
variables in a quotient of a polynomial ring by a prime ideal is transcendental
over the base field, which is done by applying the Jacobian criterion from
commutative algebra (see \cite[Lemma 19]{djs}).

\subsubsection*{Computation of the implicit equations}

As we have shown in Subsection \ref{grados}, the polynomials $P_i$'s of
Theorems \ref{despeje general dimension 0} \& \ref{despeje mejorado} and
Corollary \ref{implicita} can be interpreted as eliminating polynomials of the
image of the algebraic varieties defined in (\ref{variedades}) and
(\ref{variedades2}) under suitable linear projections. Therefore, they can be
computed by means of an algorithm  based on standard algebraic elimination
procedures (see \cite{HKPSW} and \cite{Schost}). For simplicity, we assume
$k:=\mathbb{Q}$. The following complexity result can be obtained:

\begin{proposition} \label{algoritmo}
There is a probabilistic algorithm which computes the polynomials $P_i$ of
Theorems \ref{despeje general dimension 0} \& \ref{despeje mejorado} and
 Corollary \ref{implicita} with error
probability bounded by $\varepsilon$, with $0<\varepsilon<1$, and within
complexity $O(\log(1/\varepsilon)d^2L)\ \Pi(n+m,\max_{i}\deg \mathbb{V}_i))$,
where $\Pi$ is a suitable two-variate universal polynomial. \square
\end{proposition}

We omit the proof of this result in the present article, since it is
rather long and technical, and follows closely the proof of
\cite[Proposition 46]{djs}.

\end{document}